# KOLMOGOROV'S *APORIA* AND SOLUTION
# BY
# CONSTRUCTION OF A RELATIVIZED AND *QUANTIFIED*
# CONCEPT OF FACTUAL PROBABILITY

**(A less elaborated version of this work can be found on
arXiv:0901.2301v1 [quant-ph])**

**Mioara MUGUR-SCHÄCHTER**
http://www.mugur-schachter.net/

**Abstract**. The crucial but very confidential fact is brought into evidence that – as Kolmogorov himself repeatedly claimed – the mathematical theory of probabilities cannot be applied to physical, *factual* probabilistic situations because *the factual concept of probability distribution is not defined* : it is nowhere specified how to construct factually for a given physical random phenomenon, the specific numerical distribution of probabilities to be asserted on the universe of outcomes generated by that phenomenon; nor is it known what significance to associate to the assertion of mere 'existence' of such a factual distribution of numerically defined probabilities. An *algorithm of semantic integration* of this factual numerically defined distribution of probabilities is then constructed. This algorithm, developed inside a general *method of relativized conceptualization*, involves a sort of "quantification" of the factual concept of probability. The mentioned result, while it solves Kolmogorov's aporia, fully organizes the general *classical* concept of probability, from both a factual and a syntactic standpoint. In particular, it appears that, while "randomness" can be considered to be a natural fact, the concepts of 'random phenomenon' and 'probabilistic situation' are **factual-conceptual artifacts**. As for quantum mechanical 'probabilities', it comes out – surprisingly – that, in general, they cannot be defined factually in an effective way.

## 1. INTRODUCTION

The concept of probability is ancient and intuitive. It belongs to common thinking and speaking. The mathematical formalization of this concept has begun relatively late in the history of thought (Blaise Pascal 1654) and it evolved slowly (Bernoulli 1713, Richard von Mises 1931). The first thorough axiomatic and mathematical formalization has been given in 1933 by Andreï N. Kolmogorov [1950].

Meanwhile, in physics, Ludwig Boltzmann, long before Kolmogorov's work, has introduced his famous concept of *statistical* entropy (1872-1877) which rooted the second principle of thermodynamics into atomic physics via the relative frequencies of outcome assigned to the considered 'events' [1].

Much later Shannon [1948] published his *theory of communication* (refined by Khinchin [1957]) where Kolmogorov's abstract concept of a probability measure is made use of as a given primary concept. Instead of 'events', Shannon introduced an 'alphabet' of *signs {$a_i$}, i=1,2,....n* on which he posited individual probabilities $p(a_i)$, i=1,2,....n constituting a probability 'measure' in the sense of Kolmogorov's abstract theory of probabilities. Furthermore Shannon defined as a central concept of his theory, an entropic form called *informational entropy of the source of the signs {$a_i$}, i=1,2,....n*, where in the place of Boltzmann's statistical relative frequencies of outcomes of physical events he introduced the abstract probabilities $p(a_i)$, i=1,2,....n of the signs from the alphabet *{$a_i$}, i=1,2,....n*.

For some time Shannon's concept of informational entropy seemed to permit the construction of entropic measures of complexity thus leading to a mathematical theory of complexity. But, surprisingly, 30 years after his construction of what was unanimously considered as an achieved modern mathematical theory of physical probabilistic situations, Kolmogorov became aware that his mathematical representation of the concept of probability was in fact *devoid of a factual basis* because the *factual* concept of physical probability simply is *not defined*. Therefore Kolmogorov began to claim that his mathematical theory of probabilities is not, as he had believed, an abstract reformulation of a well constructed physical concept, but exclusively an interesting mathematical construct. He also asserted that consequently his theory of probabilities cannot be made use of as a basis for Shannon's theory of communication; nor, *a fortiori*, for a concept of informational entropy that be utilizable for estimating complexities of factual entities. So he initiated another approach for measuring complexities, namely the well known theory of 'algorithmic complexity' of *sequences of signs*, which Chaitin and other authors keep developing.

---

[1] Throughout what follows the sign '....' indicates a way of saying; the sign "...." indicates a concept; and the sign «....» indicates a quotation.

In the algorithmic representation of the complexity of a sequence of signs, the semantic contents of the considered sequence of signs get entirely lost. While on the other hand, in the recent approaches on systems and on organization, the accent falls more and more heavily upon the structures of *significances*; so far however the mentioned approaches stubbornly stay purely qualitative.

So, quite confidentially, the crucial correlated concepts of factual probability, information and complex systems, are undergoing a crisis.

It is not current, nor easy, to convey both a fundamental problem and a solution of it in only a couple of pages. However precisely this is what is tried in this work.

We shall first define thoroughly the problem of factual probabilities. Then we shall construct an effective solution to this problem. This will be done inside a general method of relativized conceptualization of which the first principles have been drawn from the descriptions of microstates as these are involved in the mathematical formalism of fundamental quantum mechanics (MMS [1991], [1992A], [1992B], [1993], [2008], [2009] (MMS: M. Mugur-Schächter)).

The reader is asked to excuse a largely personal bibliography: this is not a work of erudition, it is a *construction* extracted from a work developed in extreme isolation.

## II. KOLMOGOROV'S CONFIDENTIAL APORIA:
## ABSENCE OF DEFINITION FOR A FACTUAL PROBABILITY LAW

### II.1. Kolmogorov's Classical Definition of a Probability Space

The fundamental concept of the nowadays mathematical theory of probabilities – in Kolmogorov's formulation – is a probability space $[U,\ \tau,\ p(\tau)]$ where: $U=\{e_i\}$ (with $i\in I$ and $I$ an index set) is a *universe of elementary events* $e_i$ (a set) generated by the repetition of an 'identically' reproducible procedure $\Pi$ (called also 'experiment') which, notwithstanding the posited identity between all its realizations, nevertheless brings forth elementary events $e_i$ that *vary* in general from one realization of $\Pi$ to another one; $\tau$ is an *algebra of events* built on $U$ [2], an event – let us denote it $e$ – being a subset of $U$ and being posited to have occurred each time that any elementary event $e_i$ from $e$ has occurred; $p(\tau)$ is a *probability measure* defined on the algebra of events $\tau$ [3].

A pair $(\Pi,U)$ containing an identically reproducible procedure $\Pi$ and the corresponding universe of elementary events $U$ is called a *random phenomenon*.

On a given universe $U$ one can define various algebras $\tau$ of events. So it is possible to form different associations [[random phenomenon] [a corresponding probability space]], all stemming from the same pair $(\Pi,U)$.

With respect to the previous representations of the concept of probability (Bernoulli, von Mises, etc.) – where only the 'probability law' was defined mathematically – Kolmogorov's concept of a probability *space* $[U,\tau, p(\tau)]$ has marked a huge progress: via this concept the formal representation of the factual situations qualified as 'probabilistic' became inserted in a definite way into the very elaborate mathematical syntax of the theory of measures, whereas previously these situations had been only intuitively characterized, though also numbers were made use of.

### II.2. On the physical interpretation of an abstract probability measure

The probability measure $p(\tau)$ is the unique specifically 'probabilistic' element from a Kolmogorov probability space. Now, to this very day, the application of this formal concept, to a factual situation which is unanimously considered to be 'probabilistic', has not yet been founded upon an explicit and general effective procedure for specifying the involved factual and numerically defined distribution of probabilities. It is not even clearly known what significance one has to assign to the assertion that in this or that concrete situation asserted to be probabilistic, there 'exists' a numerically defined distribution of probabilities; *a fortiori* it is not known how to identify this distribution. The specification – even only in principle – of such a procedure-and-significance would suffice for installing a concept of factual probability 'law' that be acceptable as a physical interpretation of the mathematical concept of a probability measure. This would rehabilitate Kolmogorov's mathematical representation in the status that deserves the denomination of a mathematical theory *of physical probabilities*. However the specification of such a procedure-and-significance is entirely lacking. This is surprising concerning a concept so currently utilized and which so often plays a basically important role. In this or that particular case when statistical dispersions are recorded while some set of *stable* global conditions permits to define what is called a random phenomenon, one just asserts – on the basis of symmetries – that an a priori uniformly distributed probability law on the involved universe of elementary events is justified; and in consequence of this, the probability of each event from the algebra defined on the universe of elementary events, is defined as the ratio between the number of elementary events by which the considered event can arise and the number of all the possible elementary events from the universe of elementary events. This definition, however, cannot be made use of in *any* case. For in general no obvious symmetries do come in and the a priori uniform law of the elementary events is *not* confirmed by the a posteriori effectively counted relative frequencies of the outcomes of the various elementary events from the considered universe.

---

[2] An algebra built on a set $S$ is a set of subsets of $S - S$ itself and the void set $\emptyset$ being always included – which is such that if it contains the subsets $A$ and $B$, then it also contains $A\cup B$ and $A - B$.

[3] A probability measure defined on $U$ consists of a set of *real* numbers $p(A)$, each one associated to an event $A$ from $U$ and such that: $0\leq p(A)\leq 1$, $p(U)=1$ (norm), $p(\emptyset)=0$, and $p(A\cup B)\leq p(A)+p(B)$ where the equality obtains only if $A$ and $B$ are mutually 'independent' in the sense of probabilities i.e. if they have no elementary event $e_i$ in common ($A\cap B=\emptyset$). The number $p(A)$ is defined as the value of the mathematical limit – supposed to exist – toward which any relative frequency $n(A)/N$ converges when the number $N$ of realizations of the involved repeatable procedure $\Pi$ is increased toward infinity, $n(A)$ being the number of outcomes of $A$ when $\Pi$ is repeated $N$ times.

The problem specified above still keeps quite confidential, and also vague. For the majority of physicists, for the specialists of communication, for the mathematicians who only make use of the theory of probabilities without placing it in the heart of their research, for the men in the street, a profane confidence reigns that all the important questions concerning probabilities certainly have since a long time obtained an answer in the specialized works. Beliefs of this sort arise concerning any scientific question. They are the fragile but necessary ground on which the evolution of science quietly rolls. But those who develop a research involving the foundations of the theory of probabilities are entirely conscious that today the mathematical concept of a probability measure entails a vitally important problem of interpretation. Kolmogorov ( [1963]) himself wrote (quoted in Segal [2003]) :

> « I have already expressed the view …that the basis for the applicability of the results of the mathematical theory of probability to real random phenomena must depend in some form on the *frequency concept of probability*, the unavoidable nature of which has been established by von Mises in a spirited manner…..(But) The frequency concept (of probability)[4] which has been based on the notion of limiting *frequency* as the number of trials increases to infinity, does not contribute anything to substantiate the applicability of the results of probability theory to real practical problems where we have always to deal with a finite number of trials ».

This quotation deserves much attention. One cannot be clearer. Nevertheless let us comment. At the present time there exists a more or less fuzzy but very acting belief according to which the theorem of large numbers would establish in a *deductive* way the *existence*, for any factual random phenomenon, of a factual probability 'law' of which, moreover, also the *numerical distribution* would be specified by this theorem. But *this is false*. The well known theorem of large numbers asserts only what follows (I make use of the traditional notations).

Given a set *{$e_j$}, j=1,2,....q* of events $e_j$ (or of elementary events, indifferently), *if* a probability law *{$p(e_j)$}, j≡1,2, ....q* on this set *does exist, then* for every $e_j$ and every pair *($\varepsilon, \delta$)* of two arbitrarily small real numbers, there exists an integer $N_0$ such that when the number $N$ of 'identical' reproductions of the experiment *$\Pi$* from the considered random phenomenon becomes equal to, or bigger than $N_0$, the meta-*probability*

$$P[( |n(e_j)/N - p(e_j)| ) \leq \varepsilon] \tag{1}$$

of the meta-event that [the absolute value of the difference *$(n(e_j)/N – p(e_j))$* between, on the one hand the relative frequency $n(e_j)/N$ counted for the event $e_j$, and on the other hand what is called the probability $p(e_j)$ of the event $e_j$, be *smaller* than or equal to $\varepsilon$ ], becomes *bigger* or equal to *(1-$\delta$)*. This can also be expressed in a more synthetic manner by the following well known entirely symbolic writing:

$$\forall j, \quad \forall (\varepsilon, \delta), \quad (\exists N_0 : \quad \forall (N \geq N_0)) \quad \Rightarrow \quad P[( |n(e_j)/N – p(e_j)| ) \leq \varepsilon] \geq (1 - \delta) \tag{2}$$

This same assertion is sometimes expressed less precisely by saying that *if* a probability law *{$p(e_j)$}, j=1,2,....q* does exist on the set of events *{$e_j$}, j=1,2, ...q, then* for any *j*, as $N$ 'tends toward infinity', the absolute value of the difference between the relative frequency $n(e_j)/N$ and the probability $p(e_j)$, 'tends in probability' toward *0*, i.e. it tends *nearly* certainly toward *0*. *Nearly* certainly, *not* certainly, because in the expression $\textbf{\textit{P}}[( |n(e_j)/N - p(e_j)| \leq \varepsilon ]$ the symbol $\textbf{\textit{P}}$ designates itself only a meta-*probability*, not a certainty[5].

---

[4] The brackets indicate our own specifications.

So in the theorem of large numbers the *existence* of a *factual* probability law is by no means proved. It is just *posited* that a 'probability law' does exist, without distinguishing explicitly between an abstract law that defines exclusively the general purely syntactic structure of any probability law, abstract or factual, and a factual probability law that would specify also the *numerical* distribution of the individual probabilities posited to make up this law.

What is proved indeed by the theorem of large numbers is that *if a probability law $\{p(e_j)\}$, j=1,2....q does* exist – *any* one and *non* specified numerically – *then*, as the number $N$ of the achieved trials increases 'toward infinity' (a non-effective assumption), the mathematical tendency of each relative frequency $n(e_j)/N$ of an event $e_j$ toward the numerically non specified individual probability $p(e_j)$ assigned to $e_j$ by this supposedly existing law, is itself very 'probable' *in the sense of another probability law* denoted by the symbol $\textbf{\textit{P}}$, which is also just *posited to 'exist'*.

So concerning the significance to be assigned to the assertion of 'existence' of a factual probability law, the law of large numbers offers us an infinite regression.

As for the numerical distribution of the posited probability law $\{p(e_j)\}$, j=1,2....q, a definition of it *is constructed **inside** the very theorem of large numbers* – the famous 'relative frequency definition', insured by the use of the meta-probability $\textbf{\textit{P}}$ just posited in its turn to exist – but this definition : *(a)* is *non effective* (as noted above and as stressed by Kolmogorov); and *(b)* it is constructed on the basis of the bare postulation of the existence of the two probability laws $\{p(e_j)\}$, j=1,2....q and $\textbf{\textit{P}}$, without in any way specifying in what sort of physical features of what sort of physical entity this existence consists. Indeed, in *(2)* the counted relative frequencies $n(e_j)/N$ can be conceived to play a role of specification-by-materialization of merely ideally observable limiting numerical values $p(e_j)$ that are not independently defined, *only* if they are a priori conceived to be somehow *subjected* to the postulated existence of these limits themselves: otherwise why should a convergence manifest itself?

This is the very intricately circular conceptual situation toward which point Kolmogorov's above quoted critics.

When one concentrates attention upon this circular situation it leaps to one's eyes that as long as an *independently* constructed concept of a *factual* and *effective*, *numerically specified* probability distribution, is lacking, it is improper to relate physical probabilities, with a formal system like Kolmogorov's mathematical theory of probabilities. So what is lacking is an independent and effective, factually constructed definition of the existence and of the numerical distribution of individual probabilities tied with the particular features of any given physical 'probabilistic situation'. Kolmogorov's mathematical concept of a probability measure can only be the general formal representation induced by the class of all the conceivable such effective factual definitions. It cannot be their generator. Kolmogorov's mathematical characterization of a probability measure leaves this measure devoid of any specification able to connect it with a given, singular, concrete probabilistic situation.

Confusions between the conditions to be required – specifically – concerning the description of a given factual entity, and on the other hand the conditions to be required concerning a purely syntactic framework for the representation of a whole class of descriptions of a same general type, are not rare. And quite systematically such confusions introduce long-lasting illusory problems. There exists a sort of idolatry of mathematical syntax that induces the implicit belief that it should be possible to *derive* factual data inside a syntactic system. But this is never possible.

Already before Kolmogorov, other authors also have manifested reservations with respect the applicability of Kolmogorov's theory of probabilities. For instance R. J. Solomonoff [1957] wrote[6]:

---

[5] Throughout these formulations the prefix 'meta' means that the definition of the considered event or probability *involves*, respectively, the events $e_j$ and the probabilities $p(e_j)$ and therefore it is conceptually posterior to these.

[6] The possible list of such quotations is certainly long. But this work is not a work of erudition, nor written by a specialist of the mathematical theory of probabilities. It is the work of a physicist who has been obliged to build for herself an opinion concerning the present

« Probability theory tells how to derive a new probability distribution from old probability distributions…….. It does *not* tell how to get a probability distribution from data in the real world ».

But it was Kolmogorov himself who finally developed a definitive veto concerning the applicability of his mathematical theory, to factual problems. Throughout the decade 1980 he expressed refusal of Shannon's central concept of 'informational entropy'[7]. Quite radically, Kolmogorov [1983] has advocated the *elimination* of his own formal concept of probability, from all the representations which had been considered as 'applications' of this concept. In particular, he wrote:

« **1**. Information theory must precede probability theory and not be based on it. By the very essence of this discipline, the foundations of information theory have a finite combinatorial character.

**2**. The applications of probability theory can be put on a uniform basis. It is always a matter of consequences of hypotheses about the impossibility of reducing in one way or another the complexity of the descriptions of the objects in question. Naturally this approach to the matter does not prevent the development of probability theory as a branch of mathematics being a special case of general measure theory.

**3**. The concepts of information theory as applied to infinite sequences give rise to very interesting investigations, which, without being indispensable as a basis of probability theory, can acquire a certain value in the investigation of the algorithmic side of mathematics as a whole».

## II.3. Refusal of Kolmogorov's reaction

In short, the father of the modern mathematical theory of probabilities wanted the informational problems as well as those concerning complexity, to be treated from now on without making use of the formal concept of probability. He wanted them to be treated by the means of, exclusively, combinatorial analyses of « hypotheses about the impossibility of somehow reducing in one way or another the complexity of the descriptions of the objects in question ». As for probabilities, he wanted to confine them inside the purely mathematical general measure-theory as long as no clear definition of a physical interpretation would become available. He conceived to imprison in an abstract cage the concept of probability so profoundly rooted into the concrete human experience!!!

This is a proposition made by a major thinker, so it has to be seriously taken into account. But it is an extreme proposition. Among mathematicians this proposition has been accepted without resistance and it has already changed the direction of research concerning 'complexity'. This is not surprising. For many mathematicians the physical entities are like shadows of the mathematical ones.

But for a physicist it is simply not conceivable that a formal concept like that of a probability measure – which *stems* from factuality – be unable to point in return toward an explicitly constructible factual significance. For a physicist the conceptual situation explicated above is simply scandalous.

So the problem dealt with in this work is: construct a general and effective factual definition of the numerical distribution of individual probabilities to be associated to any given 'probabilistic situation'. Thereby the assertion of *existence* of such a specifically appropriate numerical distribution of an effectively and factually defined probability law, would also acquire some definite content.

The problem formulated above will be treated here inside a general method of relativized conceptualization – *MCR* – which I keep developing since 1984 (MS [1984], [1991], [1992B], [1992C], [1993], [1995], [1997A], [2002A], [2002B], [2006], [2008], [2009]. This method has been constructed by a synthesis and an adequate generalization of results progressively obtained concerning the way in which the mathematical formalism of the fundamental quantum mechanics succeeds to signify.

---

status of achievement of the concept of probability *in the sense of physics*. So a minimum of quotations from mathematicians specialized in the abstract theory of probability measures, should suffice.

[7] The mathematical expression $H(S)=\Sigma p_i log(1/p_i)$ which possesses the same form as Boltzmann's function of *physical statistical* entropy $S=\Sigma(n(e_j)/N)log(1/(n(e_j)/N))$ but where, instead of the relative frequencies $(n(e_j)/N$ of a set of factual events $\{e_j\}$, $j=1,2, …,…q$, are inserted the probabilities from a probability measure $\{p_i\}$, $i=1,2,…q$ in the sense of Kolmogorov's theory, defined on a set of signs $\{a_i\}$, $i=1,2,…q$ emitted by a 'source of information' in order to be coded and made use of for the transmission of messages.

### III. THE FRAMEWORK FOR TREATING KOLMOGOROV'S APORIA: FROM "INFRA-QUANTUM MECHANICS" TO A "METHOD OF RELATIVIZED CONCEPTUALIZATION"

In this chapter – for the sake of self-consistency and in order to make understandable the use of the general method of relativized conceptualization – I give a telegraphic sketch of the genesis and the main features of the method of relativized conceptualization.

#### III.1. A hypothesis tied with a historical fact

There has been no equivalent, for quantum mechanics, of a Newton, a Maxwell, a Carnot, a Boltzmann, an Einstein. Quantum mechanics arose from a relatively big number of very different contributions (from Plank, Einstein, Bohr, de Broglie, Schrödinger, Heisenberg, Born, Pauli, von Neumann, Dirac, etc.) which finally led to a coherent mathematical theory of microstates – fundamental quantum mechanics – that yields predictions founded on a system of algorithms. However, up to this very day, the quantum mechanical algorithms possess a cryptic character and raise problems of interpretation. Nobody claims to fully understand how quantum mechanics manages to *signify*.

What determined these peculiar specificities?

This question, if one becomes aware of it, suggests the following hypothesis. The aim, for a human being, to construct knowledge concerning microstates, involved a cognitive situation so radically different from all those encountered before, and so *extreme*, that no individual mind has been able to dwell with it globally, in isolation. But each time that this or that physicist tried to confront the aim of constructing knowledge about microstates, this *same* very peculiar cognitive situation *acted*, without getting wholly explicit in that physicist's mind. So the construction of the quantum mechanical formalism has been orchestrated by an impersonal, very peculiar cognitive situation.

As for the way of signifying of the quantum mechanical formalism, it remained cryptic because each time that an interpretation problem was examined, over and over again the problem was much more referred to the formalism itself than to the cognitive situation which determined its structure. Correlatively, this cognitive situation and its consequences have never been characterized explicitly, thoroughly and globally.

#### III.2. A project

The above formulated hypothesis suggested a project: to make *tabula rasa* of the mathematical formalism of quantum mechanics and to try to construct, in strictly *qualitative* terms, some communicable and consensual knowledge concerning 'microstates', by obeying exclusively the constraints imposed by the involved cognitive situation and by the general human ways of conceptualizing. This project led to what I called *infra-quantum mechanics*, a sort of epistemological-physical representation of microstates, constructed *independently* of quantum mechanics but where the whole way of signifying of fundamental quantum mechanics becomes clear (MS [2008], [2009]).

#### III.3. Sketch of the construction of *infra-quantum mechanics*

What follows is summarized to the extreme. The summary is focused on the goal to bring into evidence the 'absolutely' basic source of a certain radically relativized descriptional form, and to enable the reader to grasp via its genesis the specific powers of this form, its consequences on the concept of probability, the universality hidden in it, and the potentialities involved by this universality.

Any knowledge that can be communicated without restrictions (like those involved in pointing toward, miming, etc.), is *description*. A description involves by definition an *object-entity-to-be-described* (which in general is not also an "object" in the usual sense) and qualifications of this object-entity.

The basic object-entities-of-description of quantum mechanics are what is a priori denominated 'states of microsystems' or in short *microstates*[8]. These are *hypothetical* entities from a class of which the existence is postulated beforehand on historical and methodological grounds, but which no human being could ever perceive. The construction for entities of this sort, of qualifications endowed with some kind of stability, raises difficult and deep questions. Nevertheless quantum mechanics exhibits very performing qualifications of microstates which, necessarily, are reducible to some form of description. This manifests that a *descriptional strategy* has been at work which has succeeded to overcome the epistemological difficulties. As announced, we want to explicate this descriptional strategy under the constraints imposed by, exclusively, the involved cognitive situation and the general human modes of conceptualizing.

### III.3.1. Microstates as objects of description

Consider first the entities that are the objects-of-description, the microstates. Since they cannot be perceived, it is not possible to make them available for study by just selecting them from some ensemble of pre-existing entities. Nor can one study entities of this kind by just examining observable marks spontaneously produced on macroscopic devices by admittedly 'naturally' pre-existing microstates: no criteria would then exist for deciding which mark is to be assigned to which microstate. The *unique* possible general solution has been identified to be the following one. First, to accomplish a defined and repeatable *macroscopic operation* that is just posited to generate a *given* though unknown microstate; and afterward, to try to somehow manage to 'know' something about this supposedly generated microstate.

So consider the hypothetical microstate produced by a *given* operation of generation. The goal is to acquire concerning this microstate, information cast in certain pre-established terms, namely in 'mechanical' terms involving what is called 'position', or 'momentum', or 'energy', etc. The grids for the desired sorts of qualification are conceived *beforehand* and quite *independently* of the generated object-microstate. And – with respect to *these* grids – the object-microstate emerges in general still entirely unknown, still strictly non-qualified. This assertion is not in the least weakened by the presuppositions of existence of 'microstates', in general, and of the emergence of a given sort of microstate when a given macroscopic operation of generation is realized. Even though these presuppositions insert already the generated microstate into a net of *pre*-conceptualization, so into a kind of pre-posited conceptual mould for the researched new knowledge, the particular microstate that has been generated emerges non-perceptible, so *a fortiori* still entirely non singularized *from the specific points of view expressed by the definitions of the desired qualifications* of this microstate.

But on the other hand the generated microstate emerges also *relative* – in a non removable way – to the employed operation of generation, and this permits now to *label* it: it is a result of *this*, known, macroscopically defined operation of state-generation. Let us immediately embody this possibility. We symbolize by $G$ the considered operation of generation and we subject it to the condition of being *reproducible* in a communicable way. We denote by $ms_G$ the corresponding generated microstate. Though in this incipient stage the symbols $G$ and $ms_G$ are devoid of any mathematical representation, their introduction is of utmost importance. Indeed it installs inside the realm of the *communicable*, the fact that the generated microstate, though entirely unknown from the point of view of the specific qualifications that are researched for it, is nevertheless *captured*, it is made stably available for being 'studied'. From now, by reproducing $G$, it is possible to produce as many 'copies' of the microstate denoted $ms_G$ as one wants, and each copy can be subjected to some subsequent operation of 'examination', while communicating clearly what one does, by words and signs. This, however involves a *posit*. Namely that any realization of the operation $G$ produces a replica of one and same microstate: that one, denoted $ms_G$, that is labeled by that operation $G$. In other words:

---

[8] The stable micro-*systems* themselves (electrons, protons, neutrons, etc.) have first been studied in atomic and nuclear physics where they have been characterized by specific 'particle'-constants (mass, charge, magnetic moment). *Changes* of stable micro-*systems* (creation or annihilation) are studied in nuclear physics and in field-theory. *States* of stable micro-*systems* – 'microstates' – are specifically studied in *fundamental* quantum mechanics (for Dirac the word 'state', when it is made use of concerning microscopic entities, is short for 'way of moving' (dynamics)). Inside fundamental quantum mechanics the dynamic of microstates is characterized by distributions of values of 'dynamical state-*observables*'.

A microstate can be stabilized in the role of an object-entity for qualifications via subsequent processes of examination, *if and only if* one posits *a one-to-one relation* $G \longleftrightarrow ms_G$.

The question of the acceptability of such a one-to-one relation $G \longleftrightarrow ms_G$ has been very thoroughly examined elsewhere[9]. Here we just mention that this posit simply is *unavoidable* in the considered cognitive situation. If it is not introduced one cannot *start* the desired construction of some knowledge concerning microstates. On the other hand, the consequences of the acceptance of this posit are illuminating. So we do admit it, by a ***methodological decision*** according to which:

> *That* which is obtained by any realization of the macroscopically defined operation of generation denoted $G$ – *whatever it be* – is called *'the' microstate corresponding to $G$* (the grammatical singular) and it is denoted $ms_G$.

Thereby we are now in possession of an *a-conceptual* specification – 'definition' – of an unlimited number of replicas of the object-entity *'the* microstate $ms_G$ corresponding to $G$'; namely, a purely operational-factual specification of an object-entity still strictly *nonqualified* as to its singularities inside the class a priori labeled by the word 'microstate'. Indeed $G$ is not a qualification of $ms_G$. It is *only* the specification of the way of producing it (if one knows how, say, a baby has been *produced*, this does not entail knowledge about the result of the various possible operations of qualification of that baby *itself*). But, though non-qualifying, this sort of 'definition' can be made *communicable* and consensual. This is very remarkable because it circumvents the lack of any *predicate* for defining a given particular microstate in the usual, classical way. Indeed in classical conceptualization a definition is usually realized verbally-conceptually, by the help of predicates that both define *and* qualify at the same time (open a dictionary and seek, say, 'cat'. One finds (Webster, fourth edition of the Merriam series): «carnivorous domesticated quadruped….»).

So the initial extremity of the chain of information that was to be started is now established, on the basis of *a methodological decision that separates radically, in a non-classical way, the introduction of an entity in the role of object-of-qualification, from the operations of qualification of this entity*.

### III.3.2. Qualifying a microstate: emergence of a 'primordially' statistical qualification

We can now enter upon the second stage of this investigation, the stage of construction of a certain knowledge concerning specifically the generated object-entity denoted $ms_G$.

Such as it emerges from the operation $G$, the microstate denoted $ms_G$ is not observable by man. So it has now to be brought to trigger some observable manifestations on the level of human observability. This can be realized only by use of some macroscopic apparatus able to interact with the generated microstate $ms_G$.

The interaction, however, in general *changes* the initial microstate $ms_G$.

Furthermore, the observable manifestations produced by an interaction between a replica of $ms_G$ and a macroscopic apparatus, consist of just some visible or audible marks *exhibited by the registering devices of the apparatus, not by $ms_G$ itself*. They are manifestations **transferred** upon the registering devices of the apparatus. Moreover the transferred observable marks produced by an interaction between a replica of the microstate $ms_G$ and a macroscopic apparatus, do never trigger in the observer's mind some *qualia* permitting to directly 'feel' the *nature* of the qualifying *aspect* of which the apparatus has been designed to register a qualitative or numerical 'value' (as it happens when 'red' is perceived, which is directly felt to be what is called a 'colour'). Therefore the significance of the registered transferred manifestations in terms of a given value of a given qualifying quantity, has to be entirely *constructed* in some conceptual-operational-material way. This is far from being a trivial task.

Inside the mathematical formalism of quantum mechanics – which has deliberately researched as a *mechanics* applicable to microstates – each conceptual re-definition of a mechanical qualifying quantity $X_M$ has been achieved via a formal prolongation of the mathematical definition associated to that quantity inside the classical macroscopic mechanics (which necessarily involves some connection with the classical concept of a "mobile"). But inside *infra*-quantum mechanics, mathematical representations are deliberately banished in order to bring into evidence the consequences entailed by, exclusively, the cognitive conditions and the general human ways of conceptualizing which are involved when a human being tries to construct knowledge concerning 'microstates'. Nevertheless, though in strictly qualitative terms, we want to construct a representation of knowledge concerning microstates that would be specifically comparable with the mathematical representations from quantum mechanics. So infra-quantum mechanics must somehow encompass the possibility to refer to mechanical quantities initially defined inside classical thinking and then redefined for microstates. In these conditions how can we proceed?

---

[9] Specifically for microstates in MS [2008] and [2009] and in *general* terms in MS [2002A], [2002B], [2006].

Consider a 'test'-operation $X$ that is realizable on a microstate by the use of a macroscopic apparatus $A(X)$ and each realization of which ends by a 'transfer' upon the registering devices of $A(X)$, of a set $\{\mu_X\}$ of marks[10] that can be directly perceived by the human biological sensorial systems. The set of all the transferred marks observed to emerge in this way will be called *the spectrum of **data** corresponding to the test-operation $X$.*

Let us now admit on conceptual and historical bases that what is called a 'microstate' $ms_G$ is such that for any mechanical quantity $X$ that has been defined inside the classical mechanics, there exists at least one test-operation $X(X)$ which in some sense 'corresponds' to $X$ so that it can be regarded as the transposition to microstates of the classical definition of the mechanical quantity $X$: just a posit of mere existence associated with the symbol $X(X)$, but void of any other specification. Nevertheless, on the basis of this posit the symbol $X(X)$ points now toward a previously defined mechanical quantity $X$ and it can be considered to represent a 'measurement'-interaction of which the result indicates a *numerical* value $Xj$ of $X$. So we write $X(X) \equiv M(X)$. But this, in order to be useful, must be associated with coding rule which transposes any set of observable marks $\{\mu_X\}$ produced by one realization on $ms_G$ of the test $X(X)$ into *one* definite *numerical* value $Xj$ from a set $\{Xj\}$, $j \in J$ of possible numerical values of $Xj$ assigned to the quantity $X$ tied with the test-operation $X(X)$ ($J$: an index set, here discrete and *finite* by construction, for effectiveness). If, and only if, an appropriate conceptual-operational-methodological construct is actually achieved which realizes such a coding, then the involved set of all the possible numerical values $Xj$ will be called the spectrum of the quantity $X$ attached to the test-operation $X(X)$ and $A(X)$ will be regarded as an apparatus for measuring this quantity.

Suppose that all that has been required above is insured. Then the initiated process of construction of a strictly qualitative consensual knowledge on microstates that deserve being called an infra-quantum *mechanics* can be continued, notwithstanding the fact that it consists exclusively of the descriptional consequences entailed by the constraints stemming from the cognitive situation and the human ways of conceptualizing.

Immediately however the central condition of unambiguous numerical coding of the observable data produced by the test-operation $X(X)$ raises a new obstacle: the assumption of transpositions applicable to microstates, of classical mechanical qualifications like 'position', 'momentum', 'kinetic energy', necessarily involve some *model* of a microstate, even if a very vague one. For without *any* such model – *nor any qualia* concerning the studied microstate indicated by the sets $\{\mu_X\}$ of observable data – there would be no connection whatever between the 'mechanical' description of a microstate and the classical mechanical descriptions achieved via qualifying quantities that have been extracted by abstraction from the directly observable motion of macroscopic bodies. So it would not be possible to justify why some given sort of measurement-interaction $M(X) \equiv X(X)$ corresponds to precisely this or that classical mechanical quantity $X$. (And indeed, a careful examination shows that – *contrary* to the current orthodox assertion that quantum mechanics is free from any modelization – de Broglie's initial 'corpuscular wave' model remained implicitly incorporated in the quantum mechanical mathematical algorithms which represent a measurement process) (MMS [2009A] pp. 113-118 and MMS [2009B]). This organic connection between the definability of a measurement interaction $M(X)$ and a model of microstates appears at a first sight as an insuperable obstacle inside an approach which, by the exclusivity of the accepted constraints, interdicts not only mathematical representations, but also any *specified* model attached to the general concept of microstate. But in fact this difficulty also has been circumvented. It has been transcended via a (quite non trivial) general ***frame*-*condition*** which permits to code the observed marks by the use, exclusively, of their space-time locations, thus tolerating at its core *a perfect **void** of specification of the semantic contents of the observable marks produced by the measurement-interactions M(X) tied with a test-operation X(X)* (MMS [20009] pp.118-122[11]): like in the case of the measurement interactions themselves, such semantic specifications for these results of these are only posited to exist and to be *mutually individualized* by the rule of coding accordingly to the general frame-condition.

---

[10] For the sake of effectiveness it is supposed that the number of possible distinct sets $\{\mu_X\}$ is finite (anyhow, any numerical estimation performed on these marks, even if only concerning their space-time location, introduces *units*, so discreteness).

This having been spelled out, let us ask now the following question. Can the values *Xj* which code for the groups *{µ$_X$}* of transferred observable marks produced by measurement-interactions *M(X)* of the kind just characterized above, be conceived to qualify the involved microstate *itself* ? The answer is obviously negative. Indeed the measurement-interaction *M(X)* must be conceived to *change* in general the microstate *ms$_G$* that has been initially created by the operation of generation *G*. So the observable transferred marks emerge indelibly *relative* to also this change, so relative to also the employed sort of measurement-interaction. It follows that the transferred marks characterize only *globally* the whole measurement-interaction, not *separately* the supposed object-microstate *ms$_G$*.

One can however cling to the fact that the observable marks are relative to *also* the initially created microstate *ms$_G$*. One has then to take into account that two clearly distinct processes of change of the initially produced object-microstate *ms$_G$*, corresponding to two clearly distinct measurement interactions *M(X)* and *M(X')* realized by use of two distinct apparatuses *A(X)* and *A(X')* tied with two different mechanical quantities *X* and *X'*, in general cover *two different space-time domains*. When this happens, the corresponding measurement-interactions *M(X)* and *M(X')* cannot be both simultaneously achieved for ***one*** single replica of a microstate *ms$_G$*. So – in *this* sense – these two measurement interactions are *mutually incompatible*[12]. Furthermore, a measurement evolution in general destroys the micro-*state ms$_G$* initially produced by the corresponding operation of generation *G*. It follows that if one wants to obtain for the microstate *ms$_G$* observable qualifications in terms of values of *both* quantities *X* and *X'*, in general one has to generate *more* than only one replica of *ms$_G$*, because one has to achieve the two different sorts of successions

*[(a given operation G of generation of a microstate ms$_G$),(a measurement-interaction on ms$_G$)]*

– in short *[G.M(X)]* – namely successions *[G.M(X)]* as well as successions *[G.M(X')]* (the chronometer being re-set at the same initial time-value *t$_O$* for each realization of a succession of this kind.

Furthermore, the measurement-interaction for only *one* quantity X, with a given microstate *ms$_G$*, when it is repeated via the corresponding succession *[G.M(X)]* in order to 'verify' its result, in general does *not* yield systematically one *same* value *Xj*. If this does happen for some given quantity *X*, then it does *not* happen for a quantity *X'* that is incompatible with *X* in the sense defined before. This is a fact of observation. So, in general, the results are distributed over the whole spectrum *{Xj}, j∈J* of possible values of *Xj* of the quantity *X tied with the test operation X(X)* (*J*: a discrete index set). So the global observational situation which emerges by measurement interactions with microstates is quite essentially *statistical*. And the nature of this statistical character is *primordial*[13] in the sense that it marks the *very first* knowledge that can be generated concerning microstates (MMS [2007C] (in connection with Longo [2007])), so concerning matter. Therefore – on *this* level of conceptualization *itself* – this statistical character cannot be assigned to mere ignorance of a more basic conceptualization that would have been achievable previously in individual deterministic terms, at least in principle (as it is always assumed in classical thinking concerning any sort of statistical data). Only by *models* possibly constructible some day on a higher level of conceptualization could a fully non statistical description of this or that microstate be worked out. The chronology of the levels of conceptualization *begins* with a non removable, essential, primordial statistical character.

---

[11] The ERRATUM joint to the quoted work (see the bibliography) should be taken into account.

[12] The restriction to *one* replica of the considered microstate *ms$_G$* is not explicitly required inside the current presentations of the quantum mechanical concepts of incompatibility and of complementarity, though they these concepts do involve it.

[13] 'Primordial' in the sense that it yields the *very first* observable data that can be generated concerning the studied microstate.

### III.3.3. The peculiar descriptional form tied with primordially statistical transferred qualifications

So the sort of stability that can be observed concerning a microstate – that one which can be researched on the primordial level itself of the conceptualization of microstates – can equally be *only* statistical. On this primordial level one can research a descriptional invariant only by repeating, for each given pair *(G,X)* the corresponding succession *[G.M(X)]*. If the number *N* of such repetitions is sufficiently large, each one among the possible values from the spectrum *{Xj}, j∈J* of the involved quantity *X* is realized some number of times *n(G,Xj)*, so with a relative frequency *n(G,Xj)/N*. And an observable invariant can be found only concerning the set *{n(G,Xj)/N}, j∈J* of all such relative frequencies.

But *which sort of invariant, exactly*?

The first tendency is to answer: «a *probabilistic* invariant, a *probability* law *{p(G,Xj)}, j∈J* tied with the pair *(G,X)*». But this brings us back to the problem brought into evidence in the chapter II, of the *absence* of a factual definition of the 'probability law' to be asserted in a given factual 'probabilistic situation', that be *independent* from that one – non effective and circular – that is involved in the theorem of large numbers.

We are here in presence of the *most basic* manifestation of Kolmogorov's aporia that has been formulated first inside the classical thinking.

This most basic manifestation of Kolmogorov's *aporia* might come out to *definitively* exclude any solution worked out at the level of conceptualization where quantum mechanics is formulated, because of the primordially statistical nature of the descriptions of microstates that are placed on this level[14].

So *all* that, for the moment, can be factually achieved, is what follows. To realize for each studied pair *(G,X)*, a finite number *q* of series of *N* repetitions of the corresponding succession *[G.M(X)]*, *N* taking on successively the values from some finite collection of increasingly large numbers *N1, N2,....Nk...Nq*, and to survey whether yes or not some tendency of convergence does manifest itself for the relative frequencies from the obtained sets *{n(G,Xj)/N}, j∈J, N=N1, N2,....Nk...Nq*. Nothing insures a priori the existence of such a convergence.

This existence is not a logical necessity. And if no convergence were found, one would be obliged to finally give up the aim to construct some stable observable knowledge concerning microstates. But in fact it turns out that a tendency toward convergence **does** manifest itself, for any pair *(G,X)*; a fluctuating convergence, of course, as long as the integer *Nq* is kept definite, finite, effective. In these conditions, and given the confinement inside effective procedures that has been decided here and the absence so far of any general procedure for constructing the factual probability law to be asserted in a given factual situation, one can only *substitute* some *posit* to the specification of such a law. For instance, one can posit that the relative frequencies from the set *{n(G,Xj)/N}, j∈J*, measured for the longest series of repetitions, *Nq*, of the succession *[G.M(X)]*, will be assimilated *by convention* to the unknown factual numerical distribution of individual probabilities. Which amounts to just *decide* to write *{n(G,Xj)/Nq} ≅ {p(G,Xj)}, j∈J*, i.e. to assign to the ration *n(G,Xj)/Nq, j∈J* the role played in the theorem of large numbers by what there is denoted *p(ej)*. Thereby one introduces concerning the microstate *msG* a sort of 'pre-probabilistic' knowledge founded on a mere factually observed *tendency* toward convergence, which, under cover of the dense cloud of confusion which surrounds the concept of probability, is treated *by convention* as a factual 'probabilistic' knowledge[15]. Namely, in this case, a pre-probabilistic qualification marked by *a non removable relativity to the involved triad (G,msG,M(X))*. This sort of transferred pre-probabilistic qualification, involving a conventional choice, is called here *the transferred description of the microstate $ms_G \leftrightarrow G$ via the qualifying 'transfer-view' $V(X)$* and it is denoted

$$D/G, ms_G, V(X)/$$

---

[14] Inside the mathematical quantum theory it is largely admitted more or less explicitly that the **mathematical** *formalism* permits to determine the *probability* law corresponding to *ANY factual* situation concerning a microstate. Historically, this view stems from what is called 'Born's algorithm' and possibly also from Gleason's theorem on 'probability' measures in a Hilbert space (Gleason [1957]). We will not discuss here this view which the present author does not share. We only mention it in order to submit it for discussion, noting that it is far from being unanimously regarded as an established view, as it can be found out by reading a number of works where it is variously tried to *justify* it (Destouches-Février [1946], Ballentine [1973], Deutch [1999], Anandan [20002], and others). But it might come out that, for definitive reasons of principle, a justification be conceivable *only* inside some future modelization in classical terms, of the primordial transferred quantum mechanical descriptions.

[15] After all – for now – something of this same sort is what is also systematically done in any classical probabilistic situation, more or less explicitly.

This notation reminds explicitly of the genesis of this description and the relativities which it involves ('transfer-view' $V(X)$ is just a new name and notation introduced for the measurement interaction $M(X)$). But let us stress again that the description itself – the global qualification that has been constructed – consists of *nothing more* than the partially conventional pre-probability law $\{n(G,Xj)/Nq\} \cong \{p(G,Xj)\}$, $j \in J$ introduced above.

We have explicitly noted that the strategy imposed by the cognitive situation while constructing knowledge concerning microstates has led to observable qualifications that can be posited only to *involve* this microstate, but cannot be assigned to it *alone*. This might seem to already violate the classical concept of description. So let us investigate whether at least the peculiar sort of knowledge constructed here and denoted $D/G, ms_G, V(X)/$, can be considered to be *characteristic* of the involved microstate $ms_G$, i.e. whether it can be considered to involve *exclusively* the microstate $ms_G$. Now, the answer to this question is negative: no reason can be found for asserting that the same pre-probability law $\{p(G,Xj)\}$, $j \in J$ that has been found for the studied microstate $ms_G$, so for the pair $(G,Xj)$, could never arise for also another pair $(G',X)$ with $G' \neq G$ and the same qualifying quantity $X$.

But if one considers **two** *mutually* **non** *compatible measurement-interactions* $M(X)$ and $M(X')$ achieved on different replicas of one *same* microstate $ms_G$, then it seems safe enough to consider that these act as two distinct 'directions of qualification' which together, by a sort of 'intersection', *do* determine a characterization of $ms_G$; i.e. that no other operation of generation which is different from $G$ can generate a microstate for which exactly the same pair of pre-probability laws as those obtained for $ms_G \leftrightarrow G$ with $M(X)$ and $M(X')$, does emerge. All the more so, then, if *all* the mutually non-compatible pairs $(G,X)$ are considered, where $X$ runs over all the mechanical quantities redefined for a microstate: the set of all the pre-probability laws $p(G,X)$ corresponding to all these mutually non compatible pairs can quite safely be considered to express a specificity of the studied microstate $ms_G$. So let us call it **'the'** *pre-probabilistic transferred description of the microstate $ms_G$* (mind the singular) and denote it by the symbol

$$D/G, ms_G, V/$$

where $V$ designates *'the global mechanical qualifying view defined for microstates'* consisting of the union $V = \cup V(X)$ with $X$ running over all the qualifying mechanical quantities defined for microstates. Thereby the initial descriptional form $D/G, ms_G, V(X(X))/$ which cannot be considered yet to fully characterize one given microstate, has been developed into a relativized description $D/G, ms_G, V/$ by which such a characterization is achieved.

So a transferred description of a microstate $ms_G$ consists of, exclusively, a set of one or several partially conventional *pre*-probability 'laws' on the spectrum of groups of observable marks $\{\mu_X\}$ transferred on various registering devices of various apparatuses and expressed via coding rules in terms of values $Xj$ of qualifying mechanical quantities $X$. Such a description asserts strictly *nothing* concerning how the microstate $ms_G$ 'is' itself, *nor even* **where** and **when** it 'is':

*No connected space-time support is assigned to the studied microstate $ms_G$ by its transferred description $D/G, ms_G, V/$.*

This, together with the involved coding which is stripped of any semantic content, makes the transferred description $D/G, ms_G, V/$ utterly unintelligible. Thereby it violently calls for an 'explanation', for a model of the microstate $ms_G$, in the classical sense, i.e. in terms of 'intrinsic' space-time qualifications of $ms_G$ associated with other non-transferred, intrinsic values of intrinsic qualifying aspects ('hidden parameters').

### III.3.4. The global space-time tree-like structure of the transferred description of a microstate

Let us come back now to the space-time mutual incompatibilities which exclude the simultaneous realization – on *one* single replica of the microstate $ms_G$ – of different measurement-interactions $M(X)$ that are mutually incompatible. These incompatibilities entail that the set of *all* the physical successions *[G.M(X)]* which involve one *same* operation of generation $G$, falls apart into a subset of *mutually in-compatible classes* of *mutually compatible successions [G.M(X)]*. This, by a 'geometrizing' process of integration, brings forth a pre-probabilistic whole of a *new* type, with *a tree-like space-time structure* founded on one common 'trunk' corresponding to the space-time domain $d_G(t_G\text{-}t_o)$ covered by all the realizations of the operation of generation $G$, and possessing as many measurement-interaction 'branches' as there are mutually incompatible classes of mutually compatible operations of the considered sort, each branch covering a specific space-time domain and generating on its top a corresponding Kolmogorov-type pre-probability space[16]. We shall call this structure *the pre-probability tree of the pair (G,V)* and denote it by the symbol $T(G,V)$ where *V* is the utilized view.

The fig.1 represents an example with only two branches corresponding to only two quantities re-noted for simplicity $X \equiv B$ and $X \equiv C$ and topped by, respectively, the two pre-probability spaces

$$[(C1, C2, C3,...Ck,...Cm),\ \ p(G,C)]\ \text{and}\ \ [(B1, B2, B3,....Bj,...),\ \ p(G,B)]$$

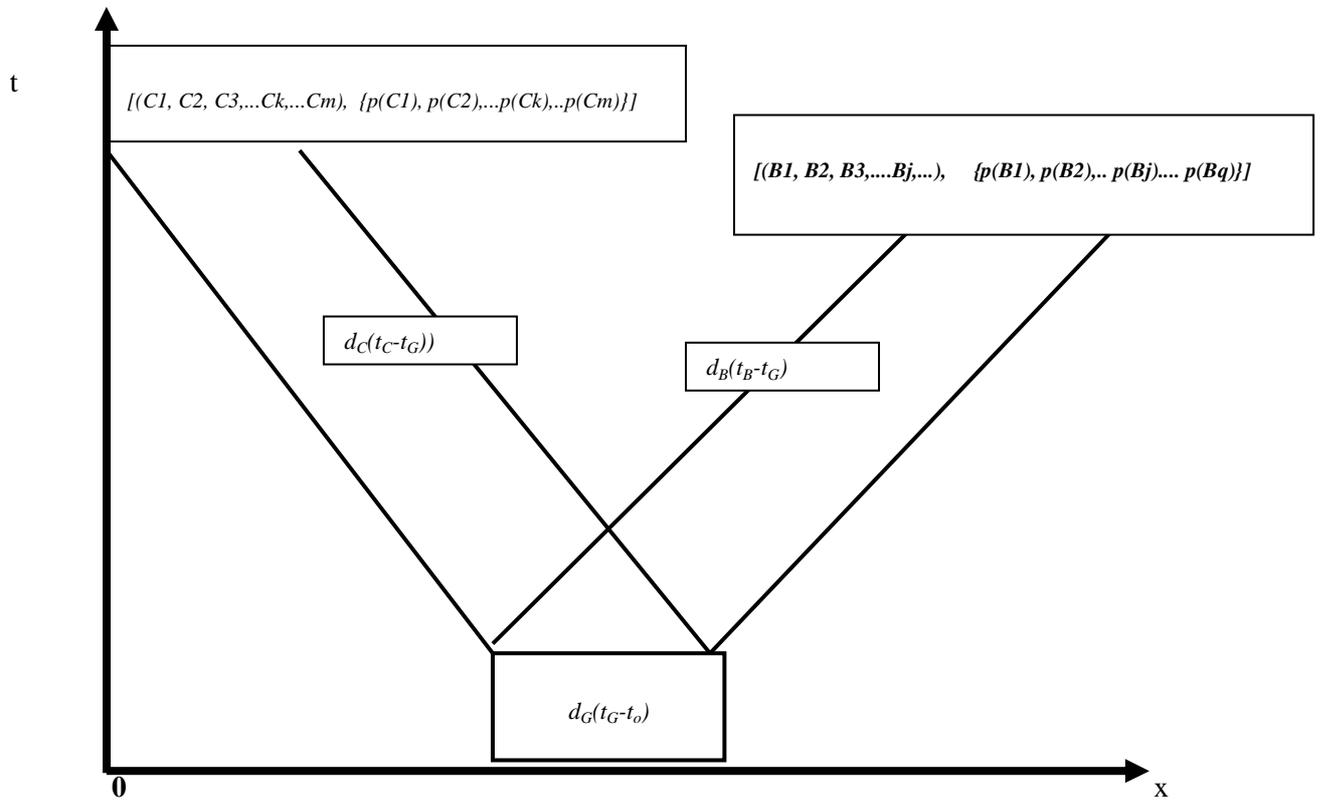

*Fig. 1*: The probability tree $T[G,(V(B)\cup V(C)]$

(The algebra on the universes of elementary events *(C1, C2, C3,...Ck,...Cm)* is skipped for the sake of simplicity and the pre-probability law *[p(G,C)]* is defined directly on the universe of elementary events; *mutatis mutandis* the same is done for the universe *(B1, B2, B3,....Bj,...)* and *p(G,B)* ).

---

[16] These space-time specifications from the 'geometrizing' integration of the *genesis* of a transferred description *D/G,msG,V/* do not in the least alter the fact that any *own, intrinsic* space-time specification of the microstate $ms_G$, is lacking.

### *III.3.5. Necessity of a deepened and extended theory of probabilities*

We have shown elsewhere that the qualitative descriptional form $D/G,ms_G,V/$ with the tree-like space-time structure $T(G,V)$ of its whole integrated, 'geometrized' genesis, introduces a number of characters which *overflow* Kolmogorov's classical concept of a probability space, quite essentially and in several important respects, namely: full *representation* of the structure of the involved random phenomenon; meta-probabilistic-dependence between the events from the mutually *incompatible* probability spaces which top the branches (accepting a specific mathematical representation not singularized before inside the general concept of correlated probabilistic spaces); pre-organized receptivity for also the logical aspects of the set of all the involved elementary events and events, whether compatible or incompatible, and which come out *not* to be expressible by a lattice structure). Thereby the concept of probability tree of the pair *(G,V)* calls for an extended and deepened theory of probabilities *unified* with a corresponding logic of all the involved events. This has been achieved in MS [1992A], [2002A], [2002B], [2006].

### *III.3.7. Conclusion on infra-quantum mechanics (IQM)*

The descriptional form $D/G,ms_G,V/$ with the geometrized, integrated tree-like space-time structure of its genesis and the consequences of it, is the heart of the strictly qualitative, physical-epistemological sort of representation of microstates constructed *independently* of the mathematical formalism of quantum mechanics and called *infra-quantum mechanics*, in short *IQM* ('infra' is to be understood here as: *beneath the mathematical formalism and partially encrypted in it*).

Here infra-quantum mechanics has been only sketched in a very simplified way. But when it is exposed in full detail it brings into light the whole way in which the quantum theory manages to *signify*. In particular, it permits to separate inside quantum mechanics that which has been introduced there by prolongation of classical models aimed toward the construction of, specifically, a *mechanics* of microstates, from what has been induced there exclusively by the cognitive situation, by the general requirements of human conceptualization, and by the aim to construct knowledge concerning microstates.

Notice however that, notwithstanding the elements of classical models encrypted in its formalism, *the descriptions of microstates from the quantum theory are **transferred** descriptions of the form $D/G,ms_G,V/$ identified inside IMQ, involving the thetree-like space-time structure $T(G,V)$.*

A systematic comparison between infra-quantum mechanics and the mathematical formalism of quantum mechanics should now permit to deal in a *unified coherent* manner with *all* the interpretation problems, and to achieve an organized dissolution of these.

But this remains exterior to the present context. The own aim of the present work is to identify the factual probability law to be asserted in any given factual probabilistic situation.

From the chapter II and from what precedes inside this chapter it appeared that the problem of defining a factual probability law in any given probabilistic situation stays open as much in the classical probabilistic thinking as in the case of the primordially statistical transferred descriptions of microstates that lie at the basis of the whole nowadays physical knowledge. And it seems worth announcing immediately that it is precisely by the use of a generalization of the descriptional form $D/G,ms_G,V/$ constructed concerning inside infra-quantum mechanics for the particular case of microstates, that we will be able to enter upon this gaping problem that vitiates the probabilistic conceptualization. In order to introduce the mentioned generalization we shall now first bring into evidence a certain universal character involved by the descriptional form $D/G,ms_G,V/$.

### **III.4. Universality and the possibility of a general method of conceptualization**

So, to achieve a description $D/G,ms_G,V/$ of a microstate it is necessary to:

**(a)** achieve the epistemic operation denoted $G$ that introduces a corresponding object-entity-of-description $ms_G$, *independently* (in general) of any epistemic action by which this entity could be qualified;

**(b)** achieve the measurement-interactions $M(X)$ that lead to qualifications of the entity $ms_G$;

*(c)* realize both operations $G$ and $M(X(X))$ in a radically creative way, by *first* generating – physically, in space-time – an object-entity-of-description that did not pre-exist (instead of just *selecting* it among already available physical objects) and *afterward* generating, physically again, observable manifestations of $ms_G$ (instead of just *detecting* properties supposed to pre-existing and to be possessed by this entity);

*(d)* realize a big number of times each succession *[G.M(X)]* for each quantity $X$ involved by the utilized view $V$, in order to try to reach – on the level of observable manifestations of $ms_G$ which irrepressibly arises with a statistical nature – an *invariant* able to constitute a qualification characteristic of $ms_G$ and endowed with an acceptably degree of stability.

The points *(a)-(d)* summarize a maximally displayed and creative way of achieving descriptions, where all the involved relativities are active and obvious and the resulting descriptional form $D/G,ms_G,V/$ is explicitly relativized to each one of the elements of the triad $(G, ms_G, V)$[17].

It is crucial to realize that such a degree of display and creativity is ignored in most of our current classical conceptualizations such as they are reflected by the natural languages as well as by classical logic, classical probabilities and classical physical theories, Einstein's relativity included. In the classical conceptualizations it has always been possible to suppose more or less implicitly that the considered object-entities-of-description *pre-exist* to the descriptional process and are 'defined' in advance by 'properties' which they 'possess' *intrinsically*, independently of any act of examination, and in an already *actualized* way. As long as the peculiar aim of describing microstates had not yet been conceived, these suppositions had never led to noticed difficulties. Therefore, classically, a description is conceived to consist exclusively in the *detection* of one or more among the properties possessed by the object-entity-of-description which itself pre-exists either as an 'object' in the usual sense, or as a 'situation', etc. The question of how an object-entity-of-description is *introduced as such* is entirely skipped. As for the process of examination that creates a qualification of this entity, it is contracted into one static act of mere detection. This last classical contraction is the source of the nowadays most explicitly stated differences between the logic and probabilities of the descriptions of microstates, and on the other hand the classical logic and probabilities. But the very deep consequences of the way in which an object-entity-of-description is generated, are quasi systematically ignored[18].

It is however noteworthy that, while in classical logic and probabilities – the two most fundamental classical syntactical structures – the descriptional form $D/G,ms_G,V/$ is not apparent, this form nevertheless is explicitly involved in many classical and quite current epistemic procedures. Indeed, once one has clearly perceived the peculiar and very difficult cognitive situation dealt with for describing microstates, as well as the descriptional strategy that permitted to dominate this difficult cognitive situation, a very paradoxical inversion arises, by a sudden variation that reminds of those which make appear certain drawings of a cube as sometimes convex and sometimes concave. What first, in $D/G,ms_G,V/$, had seemed to be fundamentally new and surprising, abruptly appears now on the contrary as endowed with a certain sort of universality, so of normality. Indeed it leaps to one's mind that:

\* any explicit and full account of a given process of description *has* to include specification of the action by which the object-entity-of-description is introduced *as such*, as well as specification of the operation, physical or abstract or both, by which a qualification is obtained for this object-entity;

\* often the above mentioned two actions are mutually independent;

\* the introduction of the object-entity-of-description is sometimes achieved by *creation* of this entity, while the operation of qualification, if it is a *physical* process, *always* – in principle at least – *changes* the object-entity, and sometimes radically, in which cases the consequences of the relativity to one or the other or both these basic epistemic actions, upon the obtained description, have to be explicitly taken into account and thoroughly analyzed.

---

[17] It might seem at a first sight that the relativity to $ms_G$ can be absorbed in that to $G$. But this cannot be done: the results of the successions *[G.M(X)]* depend explicitly on $ms_G$ and **they cannot be derived from $G$**.

[18] This is so even in fundamental quantum mechanics: there, for verbal reasons, many physicists identify erroneously the operation $G$ of generation of a microstate, with what is called 'preparation' of the microstate, which in fact is involved only in the operation of qualification of that microstate via *measurement*-interactions. This amounts to presupposing that the microsystem to be qualified is already there. In any case, the operations of generation of microstates are *not* mathematically represented inside the formalism.

For instance, think of a detective who is searching for material indications concerning a crime. What does he do? He usually focuses his attention on a convenient place from the physical reality, say the theatre of a crime, and there he first operates extraction of some samples (he cuts out fragments of cloth, he detaches a clot of coagulated blood, etc.); or he might even entirely create a test-situation involving the suspects, and insure registration of their behavior. Only afterward does he examine the gathered samples or, respectively, the behaviors registered during the test-situation.

One can equally think of a biopsy for a medical diagnosis, or an extraction of samples of rock operated by a robot on the surface of another planet, and the subsequent examinations of these entities-to-be-described. In all these cases the observer-conceptor – more or less radically – *generates* an object-entity-of-description that did not pre-exist in the desired state or quantity, in order to qualify it later by operations that are quite independent of the operation which generated these entities. And in certain cases the operation of examination so radically changes the object-entity, that, if several different examinations of this object-entity are necessary, also several replicas of it must be produced. Furthermore, the obtained qualifications arise marked indelibly by two quite distinct relativities: a relativity to the way of generating the object-entity-of-description (this way can simply exclude certain subsequent examinations), and also a relativity to the sort of examination that has been achieved. These considerations call forth the following remarks.

In the first place, the nature and realm assigned by classical thinking, to communicable knowledge, are misleadingly reduced. The whole primordial zone of conceptualization where mind *actively constructs*, out of pure factuality, the very first forms of a radically new communicable knowledge, is so deep-set, that it remained hidden beneath the two basic building blocks of all the current occidental languages, namely subjects and predicates. These do both suggest available, pre-existing elements for describing. Furthermore, the primordial, always radically creative zone of conceptualization remained cut off even from many classical *scientific* representations. Notwithstanding the well known analyzes of Husserl, Poincaré, Einstein, Piaget, and many others, not only classical logic and probabilities, but also the set theory (hence most domains of modern mathematics), modern linguistic and semiotic, etc., take their start from *language* and by use of – quasi-exclusively – *language* again. Physical operations are not considered. And factuality – via language – is widely supposed to *spontaneously* imprint, upon *passively* receptive minds, 'information' concerning already existing and actual properties of pre-existing objects. The *active* role, when it does come in, is assigned quasi exclusively to the exterior factuality, not to the mind. This attitude, in fact, is stronger and more general concerning the object-entities-of-description (grammatical subjects) than concerning qualifications (predications). As far as I know, an attempt at an integrated and systematic representation of, both, the modalities of emergence of individual object-entities-of-description, and of qualifications of these, by *deliberate* epistemic actions, and the way in which these two sorts of descriptional elements can be integrated into communicable and regulated procedures, has never been made by another author.

Quantum mechanics, by having suggested the construction of infra-quantum mechanics which led to the identification of the relativized descriptional form $D/G,ms_G,V/$, brought forth *for the first time* the potentiality of a most deep-set *general* operational method of relativized conceptualization; a general method founded on the very first interplay of what is called mind, with an entirely unknown physical factuality and that involves explicitly all the successive 'local' descriptional *aims* to be achieved, all the physical operations and devices that come in, as well as the evolving stratum of pre-existing conceptualization.

Indeed, the descriptional form $D/G,ms_G,V/$ is paradigmatic.

This is so because it has captured in it a particular embodiment of an extreme epistemic situation that is universal. Namely the situation that arises each time that a communicable and consensual representation is researched concerning some *non* pre-existing physical entity of which – a priori – only the possibility is conceived and labeled, and which, if then effectively generated, emerges non-perceivable. In such extreme circumstances one is *compelled* to a radically active, constructive attitude, associated with a maximal decomposition of the global process. All the stages of the desired description have to be *built* out of pure physical factuality, independently of one another, each one in full depth and extension. The severity of these constraints revealed a descriptional form $D/G,ms_G,V/$ so exhaustive and explicit that any other more particular form of description *must* somehow find lodging in it. This is why this form $D/G,ms_G,V/$ possesses a universal epistemological value.

As soon as this universal epistemological value of the descriptional form *D/G,ms_G,V/* has been understood, a new aim becomes conceivable : to construct a consensual canonical method of conceptualizing endowed by construction with the capacity to exclude a priori any possibility of paradoxes and false problems. Our natural way of conceptualizing incorporates quite currently seeds of problems and paradoxes which, when they bloom, block the understanding, sometimes for a very long time. If it were possible to discern whether a given process of conceptualization – not only of deduction, but quite generally of conceptualization – is 'well formed' accordingly to some explicit consensual standard, or whether on the contrary it is flawed by some genetic malformation wherefrom false problems or paradoxes can emerge, it would become possible to indefinitely develop 'clean' conceptual structures, able to carry safely and swiftly the human mind along any path of investigation or creation. Such a canonical form should consist of precisely a most displayed and explicit descriptional structure, with maximally carved out capacities. For it has to be a void form, a mould, able to offer an available, specific, and sufficiently large location, for *each* main stage of *any* descriptional process. In this or that given description, one or more locations offered by this canonical form might remain partially or totally unutilized. But this, if it happens, would be *known*, since the form will exhibit a *labeled* void of which the ampleness can be estimated. For instance, if I say «I consider what I see just in front of my eyes and this is a red surface», by reference to the maximally complete descriptional mould *D/G,V/* it will appear that in this case the two canonically distinct descriptional actions, of generation of the object-entity-of-description, and of qualification of this entity, have coalesced in the unique act of looking-just-in-front-of-my-eyes, which *both* delimits and qualifies the object-entity. So in this case the location reserved for the stage of independent generation of an object-entity-of-description remains entirely void. It will also be possible to estimate the magnitude of only partial voids and to draw consequences. For instance, imagine the assertion «I plucked this flower, I examined its morphology with a microscope, and the result is this». Comparison with the canonical mould brings forth that this amounts to a description where the object-entity is introduced – as such – by an only partially creative action (plucking a flower) while the act of examination might only very little change the object-entity initially introduced in this way. So in this case the two distinct locations reserved in the canonical mould in view of a possibly radical creativity in both the stage of production of an object-entity-of-description and in that of qualification of this entity, are both made use of, but each one to only a very reduced degree. So we know that a classical treatment (assuming the pre-existence of the object-entity-of-description as well as its invariance with respect to the process of qualification) can be posited to produce a very good approximation to the result that would be obtained by a complete canonical treatment.

### III.5. The Method of Relativized Conceptualization

The goal of the *method of relativized conceptualization (MRC)* is to offer a structure of norms of conceptualization which exclude by construction the possibility of emergence of any false problems or paradoxes. *MRC* achieves this goal *via* a systematic relativization of each step of conceptualization. The method has been constructed starting from zero. It has been developed in a deductive way. The descriptional strategy explicated inside infra-quantum mechanics for the particular case of microstates played the role of a guide. The progressive relativizations introduced with utmost detail along the whole process that leads from a zero-point of conceptualization, to any conceptual system, no matter how complex, protect this whole process from any surreptitious insertion of false absolutes. These relativizations, on their trajectory, produce a *general* equivalent, denoted *D/G,œ_G,V/*, of the descriptional form *D/G,ms_G,V/*. The form *D/G,œ_G,V/* recurs then at all the descriptional levels, like a fractal character. So hierarchical chains of relativized descriptions can be constructed, and nets of such chains. Along the mentioned path *MRC* achieves, in particular, 'clean' reconstructions of natural logic, of syntactic systems, of the probabilistic conceptualization, of the informational conceptualization.

In what follows we introduce only an enumeration (as opposed to 'construction') of those features of *MRC* that will be made use of in the following chapter[19].

---

[19] Detailed presentations can be found on http://www.mugur-schachter.net/publications.html .

*(1) MRC is founded upon a systematic relativization of any description, to a triad (G,œ$_G$,V)* where $G$ denotes the *operation of generation* – physical, or abstract or consisting of some combination of physical and abstract operational elements – by which the object-entity is made available *as such* for being qualified ; œ$_G$ denotes the *object-entity* itself introduced by $G$ ; $V$ denotes the *view* by which the object-entity is qualified.

*(2)* Any description is denoted by the symbol *D/G,œ$_G$,V/* that points toward explicitly toward the non removable relativities to the particular triad *(G,œ$_G$,V)* that is involved.

*(3) A one-to-one relation G↔œ$_G$ is posited between the operation of generation G and the object-entity œ$_G$ introduced by G. This is* **a methodological posit**, *not a fact*. But upon very careful analysis it has been found that, concerning the descriptions of microstates, this posit imposes itself inescapably. And a very attentive subsequent examination established that – if one wants to erect a method of conceptualizing that indeed banishes a priori any insertion of false absolutes – this methodological posit also imposes itself inescapably with *full generality* inside the whole class of first-stratum descriptions of *any* nature. Precisely this inescapable character of a one-to-one relation *G↔œ$_G$* entails major conceptual consequences concerning the 'primordial' sort of statistical character brought in by first-stratum, 'transferred' descriptions ((MS [2006] pp.61-66 and 213-221, [2008], [2009]).

*(4)* Any view $V$ is endowed by definition with a strictly prescribed structure:

- A view $V$ is a *finite* set of *aspect-views Vg* where $g$ is an aspect-index.

- An aspect-view *Vg* (in short : an aspect $g$) is *a semantic dimension of qualification* (colour, weight, etc.) able to carry any *finite*[20] set of 'values' *(gk)* of the aspect $g$ which one wishes to consider (for instance for 'colour' one can choose to consider only the 'values of colour' indicated by the words 'red', 'yellow', 'green' *to each one of which is associated a sample* ; the bracket surrounding *gk* shows that this symbol functions like a *unique* index different from $g$ alone ; in a definite case the indexes $g$ and *(gk)* can be replaced by any other pair of convenient signs). An aspect-view *Vg* is defined *if and only if* are defined all the devices (instruments, apparatuses) as well as all the material or abstract operations on which is based the assertion that an examination of a given object-entity via the aspect-view *Vg*, has yielded this or that – unique and definite – value *(gk)* of $g$ (or *none*).

- A view $V$ is a finite *filter* for qualification: *with respect to aspect, or values of aspects that are not contained in it by its initially posited definition, a given view V is* **blind**, *it does not perceive them*.

- The qualifications of space and time are achieved via a very particular sort of *frame-views V(ET)* (reducible, if convenient, to only a space-frame-view *V(E)* or only a time-frame-view *V(T)*).

*(5)* Given a pair *(G,Vg)*, the two epistemic operators $G$ and *Vg* can *mutually exist*, or ***not***. If any examination by *Vg* of the object-entity œ$_G$ introduced by the object-entity generator $G$ produces one well defined result *(gk)*, then the aspect-*value (gk)* of $g$ does exist with respect to $G$, i.e. there is *mutual existence* between $G$ and *(gk)*; hence, *a fortiori*, there also is mutual existence between the *aspect g* itself and the operation of generation $G$. In *this* case the pair *(G,Vg)* constitutes *an epistemic referential*. This means that in this case, if one applies to the object-entity œ$_G$ introduced by $G$, an examination by *Vg*, so if one produces the operational succession *[G.Vg]*, then one *might* obtain a corresponding "description" of œ$_G$ via the grid for qualification consisting of the aspect-view *Vg*. This happens indeed only if by repetitions of the succession *[G.Vg]* there does emerge some *invariant* result, either an individual invariant or a 'probabilistic' one (though what that means *factually* is precisely what remains to be specified in this work).

If on the contrary, what is defined to be an examination by *Vg*, when applied to the object-entity œ$_G$, yields no definite result, then there is no mutual existence between *Vg* of œ$_G$ (œ$_G$ does not exist relatively to *Vg* and *vice versa*) (for instance, a song does not exist with respect to a grid for qualifying in terms of intensity-values of an electrical current, and *vice versa*). In this case the initial matching *(G,Vg)* has to be eliminated a posteriori as unable to generate a relative description *D/G,œ$_G$,Vg/* (non significant from a descriptional point of view).

---

[20] By construction, every counting or numerical character involved in *MRC* is finite: *MRC* is conceived as a strictly effective method. Inside *MRC* any sort of infinity can be understood only in terms of *relativized absences of a priori limitation*.

The concept of mutual existence of an operation of generation $G$ of an entity-to-be-described $æ_G$ and an aspect-view $Vg$, is the *MRC*-expression of the fact that the aspect $g$ has emerged by abstraction from a class of entities to which $æ_G$ does belong; while mutual in-existence between $æ_G$ and $Vg$ is the *MRC*-expression of the fact that the entity $æ_G$ does *not* belong to the class just specified: *the pair of concepts of mutual existence and mutual inexistence constitutes the MRC-expression of the fact that a qualification can be applied only to the entities that have participated to the (social, statistical) genesis of this qualification.*

These considerations can be extended in an obvious way to also any pair *(G,V)* where $V= \cup Vg$ contains a finite number of aspect-views $Vg$. In this case one speaks of the possibility, or not, of an epistemic referential *(G,V)*.

*(6) The space-time frame-principle* asserts what follows concerning – specifically – physical object-entities.

*Any physical object-entity does exist relatively to at least one aspect-view Vg that is different from any space-time frame-view V(ET); but it is non-existent with respect to any space-time frame view V(ET) considered alone, separately from any aspect-view Vg different from any space-time aspect ET.*

Consider then a physical object-entity $æ_G$ generated by a physical operation $G$: in consequence of the space-time frame-principle, the view $V$ from any epistemic referential *(G,V)* able to generate a description of a physical object-entity-to-be-described must include a space-time frame-aspect $V(ET)$ as well as at least one aspect-view $Vg$ different from any space-time aspect $ET$[21]

*(7)* Consider a pair *(G,Vg)* where $G$ and $Vg$ do mutually exist. So the pairing *(G,Vg) does* constitute an epistemic referential where it is possible to construct a relative description of the object-entity-to-be-describe $æ_G$ produced by $G$. Let us denote this description by $D/G,æ_G,Vg/$.

If after some number $N$ of repetitions of the succession *[G.Vg]*[22] only one and the same value *(gk)* of the aspect $g$ is systematically obtained, the corresponding relative description $D/G,æ_G,Vg/$ is said to be '*N-individual*' (an 'individual description' relatively to $N$ repetitions of *[G.Vg]*), $N$ being *finite*.

If on the contrary the obtained value *(gk)* in general varies from one realization of the succession *[G.Vg]* to another one, the corresponding relative description $D/G,æ_G,Vg/$ is *statistical*, so via a very large but *finite* number $N'$ of series of $N$ repetitions of *[G.Vg]* it can only *'(N-N')-point toward'* a probabilistic description $D/G,æ_G,Vg/$ (cf. MS [2002B] pp. 145-147, [2006] pp. 75-78). But the specification of the conditions in which a *factual* 'probabilistic' invariant associated with the epistemic referential *(G,Vg)*, does 'exist' and furthermore can be identified by some effective procedure, is precisely the aim of this work. As long as the question of an effective specification of a factual numerical probability distribution associable to a factual probabilistic situation is not yet solved, one can only speak *statistical* descriptions endowed with *'(N-N')-stability'*. (This is an illustration of the way in which inside *MCR* all the involved concepts are kept rigorously effective).

The preceding assertions can be generalized to the case that the utilized view $V$ contains more than only one aspect-view $Vg$: one has then to realize – separately, in general – repetitions of all the sequences of operations *[G.Vg]* for all the aspect-views $Vg$ fro $V$. The whole of all the final qualifications thus obtained will be denoted $D/G,æ_G,V/$. This description, if some stability of this whole manifests itself, can again be found to an individual relative description, or a statistical relative description endowed with an *'(N-N')-stability'*

Let us now add a remark of crucial importance concerning the concept of relativized description. Inside a relative description $D/G,æ_G,V/$, the 'generator', the 'object-entity-of-description, and the view, are *not* fixed entities, they are descriptional **ROLES** freely assigned by the observer-conceptor to this or that entity, accordingly to his own descriptional **aims**: *the entity which in one description holds the role of the view, in another relative description can be put in the role of object-entity-of-description or of operation of generation*. This sort of freedom – characteristic of formalized representations – is the source of the unrestricted applicability of *MRC* to any process of conceptualization that is subject to the constraint of excluding by construction any false absolute.

---

[21] One can construct an infinity of space-time frame views, via various choices of axes of reference or origins of these, or various choices of differential geometric structures of reference (like Riemann geometry, for instance), or via various choices of space units and time units.

[22] In *general*, after a succession *[G.Vg]* the replica of the object-entity $æ_G$ involved in that succession either is changed by the examination via $Vg$, or it is destroyed (absorbed in a device, etc.). So in general *repetitions* of *[G.Vg]* require repetitions of also the generation operation $G$ (creation of a new replica of $æ_G$).

*(8)* A view $V$ is by construction a union of a finite number $m$ of aspect-views $Vg$, we can write $V=\cup_g Vg$, $g=1,2...m$. Each aspect-view $Vg$ introduces a *semantic g-axis* that carries its 'values' $(gk)$, $k=1,2,...w(g)$ where $w(g)$ is by construction an integer that depends on $g$. So $V$ introduces by construction the abstract *representation space* defined by the set of these $m$ semantic $g$-axes. It follows that:

> *Any relative description $D/G,\alpha_G,V/$ consists of a cloudy structure, a 'points-form' of (gk)-value-points with g=1,2...m, k=1,2,...w(g) contained in the m-dimensional representation-space of the view V which it introduces.*

If the object-entity $\alpha_G$ is of a *physical* nature, one must add inside $V$ a 4-dimensional discreet space-time view $V(ET)$ and then the relative description $D/G,\alpha_G,V/$ becomes a cloudy structure or 'form' of *space-time-(gk)-value-points* with $g=1,2...m$, $k=1,2...w(g)$, and $x,y,z,t$, some *finite* space-time grid upon which the units of space and time impose a discrete set of possible space-time values, this whole form being contained in the *(m+4)*-dimensional representation-space now introduced by the view $V$[23].

*(9)* One can form *chains* of relativized descriptions connected via common elements in their respective object-entities $\alpha_G$, (so in their involved operations of generation $G$) or in their views $V$. Along such a chain there exists a *descriptional hierarchy* or *order* : in general the order *1* is *conventionally* assigned to the first description from that chain ; the second description connected to the first one is then of order *2* with respect to this first description (a *meta-description* [24] with respect to the first one) ; the third description is assigned the order *3* and it is a meta-description with respect to the description of order *2* and a *meta-meta*-description with respect to the first description from the chain). Etc. So in general the order of a description inside a chain is relative to the process of construction of the chain.

But consider the case of a chain of descriptions that starts with a 'basic', first-stratum, 'transferred' description of a type that generalizes the descriptional type explicated for microstates in the chapter III; namely, a description in which the operation of generation creates an object-entity that has never been examined before and of which the observable manifestations – for some *non-restricted* reason – cannot be *directly* observed (for instance, the chemical structure of a sample of rock dislocated by a robot sent on the moon which is equipped with apparatuses able to identify chemical structure and to transmit the result on a computer screen from an laboratory on earth-laboratory). In such a case:

> *The initial basic transferred description determines an **absolute** beginning of a particular process of construction of knowledge. To express this, the order 0 – in an absolute sense – is assigned to it.*

*(10)* Passage from a given description from a chain of descriptions, to the following one, is commanded by *the principle of separation PS*, in the following sense.

Each relative description $D/G,\alpha_G,V/$ is accomplished inside an epistemic referential *(G,V)* where $G$ – in consequence of the methodologically posited one-to-one relation $G \leftrightarrow \alpha_G$ – is tied to *one* object-entity $\alpha_G$ and the view $V$ consists of a *given finite* set of aspect-views $Vg$ each one of which carries a *finite* set of aspect-values $(gk)$. Furthermore the relative description $D/G,\alpha_G,V/$ is achieved via some finite number of realizations of successions *[G.Vg]*. So a relative description $D/G,\alpha_G,V/$ is by construction a finite *'cell of conceptualization'* : if *all* the aspect-views from the global view $V$ have been taken into account, and *each* one with *all* its values $gk$, and after the realization of some arbitrarily large but *finite* number of successions *[G.Vg]* achieved for *all* the aspect-views *Vg* from *V* a descriptional invariant has been found, thereby the descriptional resources from the epistemic referential *(G,V)* have been entirely **EXHAUSTED**. Then, if one wants to obtain some new knowledge not already obtained by *D/G,œ$_G$,V/*, one has to bring in *another* convenient epistemic referential *(G',V')*, different from *(G,V)* by *G'≠G* or by *V'≠V*, or by both, and to construct inside *(G',V')* the new relative description *(D/G',œ$_{G'}$,V'/)* corresponding to the new descriptional aim. Now:

> The *methodological* principle of separation *PS* requires that this new description *(D/G,œ$_G$,V/)* be always achieved by a process *explicitly and entirely **separated** from the descriptional process that led to D/G,œ$_G$,V/.*

---

[23] If all these distinctions were made inside the theories of chords, the significances involved might become clearer.

[24] In logic the verbal particle 'meta' indicates an imbedding language, so it is conceived as placed 'under' the studied language. Here, on the contrary, 'meta' is deliberately assigned the significance of *'after'-and-connected-with*.

Thereby any uncontrolled coalescence or confusion between the aims and the geneses concerning two distinct relative descriptions is systematically avoided.

*(11)* Frequently, in a chain that starts with a basic transferred description of order *0*, in the immediately subsequent description of order *1*, the first description of order *0* – as a whole – is put in the role of object-entity, in order to be qualified by a certain peculiar sort of view that assigns it 'values' of an 'aspect' of *space-time connected support*, whereby the unintelligible transferred description of order *0* becomes intelligible in the sense that it gains conformity with the space-time frame principle *(6)*. A view that generates such conformity is called an *intrinsically modeling view*. The final result of a description of the specified kind, which inside a given chain acquires some order *l*, can be *detached* from its genesis. This leaves us with a *model of classical type* – with an 'object' in the currents sense – where nothing recalls any more the initial transferred description. The behavior of such models can be represented in 'causal' terms, *it admits **a deterministic principle***.

In this way, inside *MRC* there emerges a *split* inside the pool of all the relativized descriptions achieved at any given time. Namely, the very first, basic, transferred relative descriptions of absolute order *0* from this pool constitute *a primordial stratum of conceptualization*. And the corresponding classical models of the transferred descriptions from the primordial stratum, together with the progressively more and more complex forms acquired by them or with their insertions in nets of more complex conceptual structures, constitute an evolving classical stratum of conceptualization of indefinitely growing thickness.

Thereby *MRC* incorporates the famous *'[quantum-classic] cut'* and explains it inside a generalization in terms of a concept of a ***universal** transition '[(transferred descriptions)-(classical descriptions)]'*.

*(12)* According to *MRC* any *knowledge* that can be communicated in a *non* restricted way, is *description* (the action of 'pointing toward' restricts to real or virtual co-presence inside some delimited space-time domain, so do also mimics, emotional sounds, etc.). Only descriptions can be *unrestrictedly* communicable *knowledge*, not 'facts' which are exterior to any psyche, nor psychic facts (emotions, desires, etc.) which are not expressed by some more or less explicit description, verbal or of some other constitution. When we say « I know this house » we spell out an illusion, because of unawareness or only for the sake of brevity; only the assertion « I know the *descriptions* (plural) of this house » would correctly express the situation toward which we want to point.

*(13)* When the concept of probability is re-constructed inside *MRC, the elementary events and the events from a probability space acquire the conceptual status of relativized **descriptions**:* their *MR*C descriptional status (role) *ceases* being only that of object-entities-"of-description" *œ$_G$*; it becomes that of relative *descriptions* of object-entities *œ$_G$*. This has come out to be a quite essential progress that entails many clarifying consequences as well as a considerable increase of the power of discrimination and expression.

This mere enumeration of the main features from only the kernel of *MRC*, should suffice for finally entering now upon the specific problem raised in this work.

# IV. CONSTRUCTION INSIDE *MCR* OF
# AN EFFECTIVE PROCEDURE FOR DETERMINING
# FACTUAL NUMERICAL DISTRIBUTIONS OF INDIVIDUAL PROBABILITIES

## IV.1. Games with a parceled painting

Throughout this chapter IV we place ourselves inside the *classical* stratum of conceptualization. So the imagined object-entities involve 'objects' in the classical sense (dice, apparatuses, tables, vehicles, roads, and so forth).

This preliminary investigation will consist of a succession of examples. These will familiarize with the use of *MRC*.

By passage from a small obviousness to another small obviousness there finally will emerge a novelty: the definition of an effective procedure permitting to research and – in general – to identify an open set of relativized factual probability laws that can be asserted in any given factual probabilistic situation.

### IV.1.1. Relativized parceling and notations

Consider the puzzle of a painting $P$ representing a landscape, containing *100* square pieces $\sigma$. Consider also a spatial grid that can be superposed to the integrated solution of the puzzle of $P$. On this grid each square $\sigma$ is localized by the specification of its two space coordinates $(x_k, y_h)$ where $x_k$ is an element from a set of *10* successive equidistant coordinates *{$x_k$}, k≡1,2...10* marked on a horizontal space axis *ox* superposed to the lower edge of the painting $P$, while $y_h$ is an element from a set of *10* successive equidistant coordinates *{$y_h$}, h=1,2...10* marked on a vertical space axis *oy* superposed to the vertical left edge of $P$. The label $(x_1, y_1)$ indicates the square from the left lower corner of $P$ and the left lower corner of this square is the origin *0* of the plane Cartesian system of reference *xoy* attached to the grid superposed to $P$ ; while the pair $(x_{10}, y_{10})$ indicates the square from the right upper corner of $P$.

Consider an epistemic referential *($G_P$,V)* where : the object-entity generator $G_P$ is a 'selector' that selects as an object-entity the integrated solution of the puzzle of the painting $P$; $V$ is a view which consists of three aspect-views defined as follows :

* A space-frame-view defined as the union $V(E)≡V(El) \cup V(E\phi)$ where : $V(El)$ is a frame-view of *spatial location* (*l* : location) of which the possible values are the *100* pairs of spatial coordinates $(x_k, y_h)$, k=1,2,...10,  h=1,2,...10  (so a square $\sigma$ examined via the view $V(El)$ leads to the description $D/G_\sigma, V(El)/$ of spatial location of $\sigma$ consisting of one among the pairs of coordinates $(x_k, y_h)$, k=1,2,...10, h=1,2,...10) ; $V(E\phi)$ is a frame-view of *spatial form* endowed with a very big number of 'values of form' (this amounts to the introduction of a very small unit of length that permits to reproduce satisfactorily any perceivable contour).

* A colour aspect-view $Vc$ endowed with a set of colour-values rich enough for insuring that a relative description $D/G_\sigma, \sigma, Vc \cup V(E\phi)/$ yields a form-of-colour which reproduces 'satisfactorily' that one perceived on $\sigma$ by a normal human eye. (However, since everything in the definition of any view is by construction discrete and finite while any view acts like a filter, the total number of possible distinct 'values-of-colour-form' is discrete and finite). The view $Vc \cup V(E\phi)$ can be synthetically rewritten as $Vc \cup V(E\phi) ≡ Vc\phi$ where $V\phi$ is a *view of colour-form*.

With these definitions and notations the description of the integrated puzzle of the painting $P$ achieved inside the epistemic referential *($G_P$,V)*, has to be written as $D/G_P, P, V(El) \cup Vc\phi$ /.

Consider now a 'local' epistemic referential *($G_\sigma$,V)* where $G_\sigma$ selects as object-entity only *one* square $\sigma$ while the view $V$ is the same one is in the referential *($G_P$,V)*. Then a relative description corresponding to *($G_\sigma$,V)* is to be written as $D/G_\sigma, \sigma, V(El) \cup Vc\phi/$ : it consists of a 'colour-form' covering the selected square $\sigma$ and which is located as indicated by its 'value' $(x_k, y_h)$ of spatial location.

Let *Vac* be a new aspect-view of *'approximate-colour'* endowed with *q uniform* approximate-colour values *j, j=1,2,...q* (this square is approximately of this uniform shed of red, that square is approximately of this uniform shed of blue, etc.).

If in the local relative description $D/G_\sigma, \sigma, V(El) \cup Vc\phi/$ of a square $\sigma$, the space-location aspect-view $V(El)$ is suppressed one obtains a local description $D/G_\sigma, \sigma, Vc\phi/$ of a square $\sigma$ achieved inside the referential $(G_\sigma, V c\phi)$ where any *direct* indication of spatial location is filtered out.

If furthermore, in this new local relative description $D/G_\sigma, \sigma, Vc\phi/$, the view of colour-form $Vc\phi$ is replaced by the view *Vac* of uniform approximate-colour, the value of colour-*form* that covered the considered square $\sigma$ is equally filtered out and a new relative description achieved inside the referential $(G_\sigma, Vac)$ arises – to be written as $D/G_\sigma, \sigma, Vac/$ – where, in consequence of the *uniformity* of the approximate-colour values assigned to *Vac*, one looses now also the perceptibility of any affinity or repulsion between the form-of-colour reaching a border of the considered square $\sigma$, with respect to another form-of-colour reaching another border of another square $\sigma$. So one ceases to be able to play puzzle with the *100* squares described via exclusively *Vac* : this time any hint of some connection between the considered square $\sigma$ and the global 'significance' carried by the integrated painting $P$, is lost.

Suppose now that the global dimensions of the picture $P$ and the distance between two successive values of the $x_k$ or $y_h$ coordinates, are *such* that:

*(a)* Any square $\sigma$ is small enough for carrying only *one* approximate-colour-value $j$. Then its relative description $D/G_\sigma, \sigma, Vac/$ via the view *Vac* of uniform approximate-colour entirely consists of only one uniform approximate-colour : it reduces to just its unique approximate-colour value $j$. So we can write $D/G_\sigma, \sigma, Vac/ \equiv Dj$, $j=1,2,....q$. Then we have $\{D/G_\sigma, \sigma, Vac/\} \equiv \{Dj\} \equiv \{j\}$, $j=1,2,....q$.

*(b)* Any given partial description $Dj \equiv j$ is realized on *much* more than only one square from $P$. Thereby, by construction, the cardinal $q$ of the set of *mutually different* relative descriptions $\{Dj\} \equiv \{j\}$, $j=1,2,....q$ is *much smaller than 100*.

So finally, each one among the *100* squares $\sigma$ can be taken knowledge of via three distinct views : the frame-view $V(El)$ of spatial location , the frame-view view of $Vc\phi$ of colour-form, and the view *Vac* of uniform approximate-colour.

Let us now mix the squares and throw them all into a ballot box.

Starting from this point we define a succession of 'games' which will lead to the announced interesting conclusion.

### IV.1.2. Game illustrating the power of reconstruction contained in space (or space-time) order

Let us accomplish the *100* possible successive extractions of a square $\sigma$ from the ballot box and look at each extracted square via the frame-view $V(El)$ of spatial location. This, for each square, yields a description which places each square at the place, on the reference grid subtended by axes *xoy*, which is indicated by the obtained coordinates $(x_k, y_h)$. Since any view acts as a filter, this happens *without having taken into account the colour-form carried by it, nor the uniform approximate-colour $j$ defined on it by the view Vac*. Nevertheless, after exactly *100* extractions, the global painting $P$ is reconstructed. Though the order of extraction of the squares will have been random, each individual act of progression toward the reconstruction of the global painting $P$ will have been accomplished in a way marked by *certainty*, while the global process will have been *finite*: *the spatial grid of reference possesses a power of topological organization which is independent of the 'semantic content' of the squares*.

These remarks extend in an obvious way to the case also of a space-time grid.

### IV.1.3. Puzzle with only one replica of P

Let us now proceed differently. Let us again mix the squares and shed them into the ballot box. Let us then make again the *100* possible successive extractions of a square from the ballot box. But this time, let us make use of – exclusively – the view $Vc\phi$ of colour-form. So each square is perceived by its relative description $D/G_\sigma, \sigma, Vc\phi/$. The *label of space location $(x_k, y_h)$* as well as well as its label $j$ of uniform approximate-colour are filtered out, they are ignored. In these conditions, again, after exactly the *100th* extraction the global painting $P$ will be reconstructed. But in general, for finding the right place where to put an extracted square, we will have had to fumble around by trials and errors; but, guided by the structure of the form-of-colour carried by the square, we will have finally identified the 'good' place of the square. And the structure of the form-of-colour of the square will have been useful mainly by its content in the proximity of the *borders* of the square where, for each given border, it determines a sort of *neighborhood-coherence* with the form-of-colour reaching a *unique* other border of another square. A sort of attraction by semantic continuity acts between the two mentioned borders and, on the contrary, a sort of repulsion by semantic discontinuity works between the form-of-colour that reaches the initially considered border of the considered square, and any other form-of-colour reaching any border of any other square. This time the independent power of topological organization of the space coordinates will have been filtered out and replaced by these 'attractions via semantic continuity' or 'repulsions via semantic discontinuity'. And again nothing infinite will have been involved and nothing will have been random – if abstraction is made of the randomness in the order of extraction of the squares – notwithstanding the presence of trials and errors. For, quite obviously, the trials and errors are tied with features of the defined situation, of which the nature is radically different from that of a predictive uncertainty in the probabilistic sense.

This example, like the preceding one, can be extended in an obvious way to the case of an 'evolving picture' fragmented in space-*time* cubes of which the space-time labels are by-passed, while exclusively other descriptional contents are considered, with the attractions by continuity on the borders and the repulsions by discontinuity on the borders which these contents entail. (During the research of a criminal, for instance, in essence, one plays a generalized space-time puzzle game).

### IV.1.4. Puzzle with several replicas of the painting P

Let us now provide ourselves with *1000* replicas of the fragmented painting *P* and let us proceed with these in the same way as we did above for one replica : we mix together all the *100 000* squares which we now possess, we shed them all into the pool-box, and then we extract the squares one by one, ignoring the space label and the approximate-colour label imprinted on it and searching for the square an appropriate place on one or other among the *1000* void space-time grids placed in front of us. What will happen? After *100 000* extractions from the ballot-box we *certainly* shall have entirely reconstructed all the *1000* replicas of the fragmented painting *P*. But this will have been achieved only after quite a lot of trials and errors and not by a neatly separate completion of the replicas, in succession, but by an intermingled process of completion of all the replicas, involving leaps from one replica to another one. In general, only by the last extractions will all the *1000* replicas have entirely separated from one another.

In principle, no essentially new features are brought in by the use, instead of *1000* replicas, of $10^n$ replicas with *n* an arbitrary whole number. And this game also can obviously be extended to a set of 'evolving paintings'. And again nothing infinite will have been involved and – notwithstanding the presence of trials and errors – nothing will have been random if abstraction is made of the randomness in the order of extraction of the squares.

A puzzle game, no matter how complex and big, involves randomness exclusively in the order of extraction of the squares.

*The attractions by semantic continuity on the borders of the squares and the repulsions by violation of semantic continuity on the borders, exclude randomness from the final reconstruction of any number of replicas of the global entity which has been parceled.*

### IV.1.5. Probability game with one replica of the painting P

How, then, does 'probabilistic randomness' emerge? By a modification which, at a first sight, will seem insignificant, suddenly all the characters of a 'probabilistic situation' will come in: unending sequences of elementary events, the corresponding statistical relative frequencies, probabilistic randomness and probabilistic convergence. The announced apparently insignificant modification will reveal itself to have been in fact a radical conceptual jump.

We make use of the same parceled painting *P* involved in the preceding paragraphs. But this time, instead of a puzzle game, let us play, with just one replica of this puzzle, the following 'probabilistic game'. Let us mix the squares and shed them into the ballot box. Then let us extract a square. Let us use exclusively the view *Vac* of uniform approximate colour, note the value of the index *j* that appears in the corresponding relative description *Dj, j=1, 2....q*, and then *drop the examined square back into the ballot box* (both the aspect *Vφc* of colour-form and the space frame-aspect *V(El)* of spatial location remain dumb, *so a fortiori* the semantic continuity on the borders of the square remain inactive). Let us then mix the squares from the ballot box and repeat the same procedure an arbitrarily big number of times.

I assert that this time, in consequence of the specified modification of the procedure, we find ourselves in a standard 'probabilistic situation'. Indeed, contrary to what happened in all the preceding cases, this time, *before* each extraction, a certain set of *invariant* conditions is reconstituted, which defines – in the usual sense of probabilistic language – a 'reproducible procedure' *Π* and a *stable* universe of elementary events $U≡\{j\}$, $j=1,2....q$, so a random phenomenon *(Π,U)*. Since according to *MRC* (point *(9)*) any communicable knowledge is description, let us rewrite explicitly the universe *U* as a universe of relative descriptions $U≡\{Dj\}$, $j=1,2....q$.

What will happen in these new probabilistic conditions? *Can that be predicted?*

If the number of successive extractions and droppings back into the ballot box is very much bigger than the number *q* of elements from *U*, one can make the two following rather obvious remarks *R1* and *R2*.

**R1**. Since the whole initial content of the ballot box is reconstituted after each drawing, all the descriptional values *j=1,2,....q* that had been possible before some given extraction, are equally possible for the following one. From one extraction to another one, no possibility is irreversibly 'consumed', as it happened in the cases considered in the preceding paragraphs. This entails stability of the global factual situation.

**R2**. Correlatively, the content of the ballot box is never exhausted. Nothing brings any more to an end the sequence of results which can be obtained by repetition of extraction-dropping. This sequence is of arbitrary length, it can increase 'toward infinity'. This is the source of potential non effectiveness.

I add now two other assertions which are not perceived as certainties.

The first one is the answer *A1* to the following question *Q1*: "If one continues indefinitely the repetitions of extraction-dropping, will all the *q* values of the index *j* of medium colour show up, or not ?"

The second assertion *A2* is the answer to the following question *Q2*: "If the repetitions of extraction-dropping are continued indefinitely, how will evolve the relative frequency $n(j)/N$ of the outcomes of a given value *j* of the index of medium colour ?".

I now assert a *psychological fact*: after a short reflection, the following answers *A1* and *A2* to the questions *Q1* and *Q2* will gain quasi unanimous consensus among persons skilled in current probabilistic thinking.

**A1**. It is *nearly certain* that if the number *N* of repetitions of extracting-dropping becomes big enough, all the *q* values of the index *j* of medium colour will show up.

**A2**. If the number *N* of repetitions of extracting-dropping is increased without a priori limitation, then – earlier or later but *nearly certainly* and *for any j* – the relative frequency $n(j)/N$ of the outcomes of a given value *j* of the index of uniform approximate-colour will manifest a certain convergence. Namely, the value of *the relative frequency $n(j)/N$ will tend to reproduce the value of the ratio $n_P(j)/100$ which refers the number $n_P(j)$ of squares from the integrated puzzle of the painting P which carry the considered value j of approximate-colour, to the total number100 of all the squares from the puzzle of P.*

But *why* should there be a convergence? And quite especially, *why precisely toward this ratio $n_P(j)/100$ defined on P*? And why, in both formulations *A1* and *A2*, should one assert a 'nearly certainty' instead of, clearly, a certainty? Because, more or less explicitly, in the minds of those who adhere to the answers *A1* and *A2*, some equivalent of the following reasoning takes place.

"Since after every extraction and registering of the obtained *j*-value, the square is released back into the ballot box, and since the extractions are allowed to be repeated indefinitely, there exists no basis for strictly excluding a priori, concerning a sequence of arbitrarily big length *N*, the outcome of any one among the different possibilities *{j≡1,2,...q}* ; nor, moreover, the outcome of any order of succession of *j*-values from *{j=1,2,...q}* ; nor the outcome of any one among all the global statistical distributions *{n(j)/N, j=1,2,...q}*, $\Sigma_j n(j)/N=1$, of relative frequencies *n(j)/N* that are constructible for a given *N*, with *j*-values belonging to the universe *{Dj}*, *j≡1,2,...q* of relative descriptions. In the conditions of indefinite repeatability that have been posited here, any outcome of any feature that cannot be a priori excluded on the basis of some specified reason, has to be a priori admitted as possible. These two formulations have the same significance, so any distinction between them would amount to a contradiction. For instance, nothing permits to strictly exclude a priori the maximally unbalanced statistical distribution which, for any given *N* and *j'*, is characterized by *n(j')/N=1*, i.e. *n(j')=N*, *n(j)=0* for any *j≠j'* (with *j'=2*: *2222222222..... N* times). Indeed if in the first extraction it has been possible to find a square carrying *j=2*, since that square has been released back in the ballot box before the second extraction, the same possibility holds also for the second extraction, and so on, indefinitely. But nothing excludes either to find *j≠2*. This entails the answer *A1*.

However we know that the number of squares in the pool box and the number of possible approximate-colour values *j* are both finite and that *any square comes from the puzzle of the integrated painting P*. In these conditions, *before* each extraction it is natural to expect more to find on the extracted square a *j*-value of approximate-colour which, on the integrated painting *P*, is repeated, say, on *10* different squares, rather than to find the *j*-value of approximate-colour which on the integrated *P* is repeated, say, on only *2* different squares. What is effectively found *after* an extraction leaves invariant the reasonableness of the specified expectation before that extraction. We must avoid confusion between a priori and a posteriori as well as between 'possible' and 'probable'. So, since we know that before each extraction the pool box contains *exclusively* the *100* squares which compose the total number, let us denote it $n_{PT}=100$, of puzzle-pieces from of one replica of the painting *P*, it is natural to expect a priori that in a sufficiently long sequence of *j*-results each possible *j*-value be obtained a number of times approximately proportional to the number of squares on which this j-value is realized *on the integrated painting P*; and to also expect that while the number *N* of accomplished extractions increases, the relative frequency *n(j)/N* be found to tend to converge, for each given *j*-value, *toward the ratio $n_P(j)/n_{PT}=n_P(j)/100$ realized for that j-value on one integrated replica of the painting P*. In the posited conditions, any different assumption would be devoid of support, while this one – in a certain sense – simply *follows*.

Indeed the global form of one replica of *P* is contained in there, the pool box, even if it is parceled. So, in the long term, it *must* manifest itself via any view that is not entirely blind with respect to it. Now in the conditions of our probability game, the unique active view is the approximate-colour view *Vac* endowed with the possible values *j≡1,2,...q*. This view is not entirely blind with respect to the form of *P*. And in the conditions of our probabilistic game, the *unique* possible manifestation of the global colour-form aspect of *P* that is possible via the approximate-colour view *Va*, consists of a set of relative frequencies *{n(j)/N}, j=1,2,...q* which reproduces the set of ratios *{n_P(j)/n_{PT} }, j=1,2,...q* from the global colour-form aspect of *P*. So such a set of relative frequencies is what has to be expected, by default. This amounts to the asserted convergence.

However this has to be expected nearly with certainty, not with certainty. This is entailed by the conditions which ourselves have posited: these conditions simply *exclude* the assertion that each relative frequency *n(j)/N]* will certainly converge toward the corresponding ratio *{n_P(j)/n_{PT}= n_P(j)/100*, so all the more that it will strictly reproduce this ratio.

Indeed it has already been pointed out that any sequence of *N* results *j* is possible, even a sequence *kkkkkkkkk.....*of *N* results *j=k*. But we are reasoning inside the abstract framework of the concept of probability[25], so for the probabilities on a universe of events of any sort, there is a condition of *norm* : the sum of all the probabilities assigned to the events from the considered universe, whatever these be, must be equal to *1*. Consider than a sequence $\sigma_\omega(N,j)$ of *N* results *j* where '*ω*' is an index of *statistical structure {n(j)/N}, j≡1,2,...q* and *N* is any whole number. For any give *N* there exists a corresponding *finite* set *{$\sigma_\omega(N,j)$}, j≡1,2,...q, ω=1,2,......v*, of *v* mutually distinct statistical structures constructible with *N*. These constitute a universe of events (meta-events, with respect to the events from *{j, j=1,2,...q}*). And the probabilities *p($\sigma_\omega(N,j)$), ω=1,2,......v* assigned to these new (meta)events are also subject to the condition $\Sigma_\omega \, p(\sigma_\omega(N,j))=1$. So any sequence $\sigma_\omega(N,j)$, while on the one hand it is possible a priori, on the other hand it 'consumes' inside the condition $\Sigma_\omega \, p(\sigma_\omega(N,j))=1$ a certain 'quantity of probability'. This interdicts to assign a priori certainty (probability *1*) to any given sequence $\sigma_\omega(N,j)$ : if one did this, thereby, contrary to the initial assumption of a priori possibility of *any* sequence, he would a priori exclude – for the considered *N*, but whatever it be – the possibility of all the sequences *(σ_ω(N,j))* but only one among them. This would be a contradiction. So a certain and strict convergence toward *all* the ratios $n_P(j)/100$, is excluded by the very rules of our probabilistic game.

---

[25] It is **not** circular to introduce such considerations: here only the concept of a *FACTUAL* probability law is acknowledged to be still undefined, but the abstract mathematical probabilistic syntax introduced by Kolmogorov is *accepted*, at least as an initial basis.

But *nothing*, in the rules of the probability game, does interdict the intuitive notion that with sufficiently large numbers *N each* relative frequency *n(j)/N* would – nearly certainly – come arbitrarily near to the corresponding ratio *{$n_P(j)/n_{PT} = n_P(j)/100$*. This is precisely the answer *A2* to the question *Q2*.

So the quasi intuitive motivations which underlie the answers *A1* and *A2* to, respectively, the questions *Q1* and *Q2*, are now explicit. These, of course, are not motivations exiting inside any mind. They are trained motivations generated by, precisely, a deep understanding of the theorem of large numbers. So let us compare their manifestation concerning the special case of the probability game with the puzzle of the painting *P*, with the general theorem of large numbers.

### IV.2. An effective factual probability law in the case of the 'probability game' with the painting *P*

At a first sight, the motivation brought forth above for the answers to the questions *Q1* and *Q2*, seems trivial. But in fact it discloses a conclusion which, itself, is far from being trivial. Indeed from *A1* and *A2* there finally emerges – for the particular case of a probability game with the picture *P* – an *effective* definition founded upon 'real facts', of the so elusive concept of a factual probability law. And it is by reference to the law of large numbers that this definition imposes itself. For when one writes

$$\forall j, \quad \forall (\varepsilon, \, \delta), \qquad \exists N_0: \quad \forall (N \geq N_0) \; \Rightarrow \; \boldsymbol{P}[(\, |n(e_j)/N - p(e_j)\,|) \leq \varepsilon \,] \geq (1 - \delta) \qquad (2)$$

via a mere *identification of terms* one is led to clearly perceive that the expression *(2)* of the theorem of large numbers can be regarded as a rigorous mathematical translation of precisely the partially intuitive and partially 'reasoned' answers *A1* and *A2*. Indeed let us set

$$e_j \equiv Dj, \qquad \{p(Dj)\} \equiv \{n_P(j)/n_{PT}\} \equiv \{n_P(j)/100\}, \quad j=1,2...q, \qquad (3)$$

This yields the form *(2)* where the numbers *{$n_P(j)/100$)}, j=1,2...q*, satisfy all the conditions to be imposed upon a probability law (cf. the note 3), whether formal or factual (they are real positive numbers – here *rational* numbers – they obey the norm condition $\Sigma_j n_P(j)/100)=1$, etc.).

So, in the case of the probability game with the picture *P* the set of ratios *{$n_P(j)/n_{PT}$} ≡ {$n_P(j)/100$)}, j=1,2...q* defines on the set of events *{Dj}, j≡1,2...q* a quite definite and *effective* factual probability law

$$\{p_F(Dj)\} \equiv \{n_P(j)/n_{PT}\} \equiv \{n_P \,(j)/100\}, \quad j=1,2,....q, \qquad F: factual \qquad (4)$$

It is striking to notice that the definition *(4)* amounts to make use of the intuitive concept of the probability of an 'outcome' as [the number of 'favorable cases']/[the number of all the possible 'cases']:

> The theorem of large numbers, established inside the mathematical theory of measures, implies this intuitive definition.

But such as it is implied by the theorem of large numbers, this intuitive definition is *non effective* and *absolute*. In this sense, so such that it is involved in this theorem, it is banished by *MRC*. On the other hand the construction that led to the definition *(4)* has been drawn into evidence by the use of the descriptional *relativities* imposed by *MRC*. This remarkable agreement-and-dissention between the theorem of large numbers and *MRC* might conceal a clue in the search of elucidation of the connection between factuality and syntax in the case of the concept of probability.

So, in the particular case of the probability game with the puzzle of the painting *P*, the problem of the construction of a factual probability law constructed "on the basis of real physical facts", has found a solution: Starting from the integrated painting *P*, by a 'probabilization' involving a puzzle founded on *P* and a 'simplification' of the descriptions of the pieces of this puzzle by passage from the initial elements $D/G_{c_i}\sigma(x_k,y_h),Vc \cup \mathcal{N}\phi c \mathcal{N}(E)/$ of this puzzle, to the universe of events *{Dj}≡{ j },* *j=1,2,....q,* we have finally constructed a standard factual probability space in the sense of Kolmogorov

$$[\{Dj\},\ \tau_X,\ \{n_P(j)/n_{PT}\}],\quad j=1,2,....q \qquad (5)$$

where $\tau_X$ denotes *any* algebra on *{Dj}, j=1,2...q,* while the factual probability law *{n_P(j)/n_{PT}},* *j=1,2,....q* is defined *directly* on the basic universe *{Dj}, j=1,2...q* (which then determines also the factual probability law on $\tau_X$, whatever its specification)[26].

This conclusion, together with the questions *Q1, Q2* and the answers *A1, A2* which led to it, involve a definite solution to the question, also, of the *significance* to be assigned – in *this* case – to the assertion of merely the *existence* of a factual probability law. Indeed in the answers *A1* and *A2*, the belief in the existence of a factual probability law has been founded upon the fact that before each extraction of a square, a *whole* replica of the parceled painting *P* was contained in the pool box, and nothing else. So the significance of the 'existence' of a factual probability law is [the systematically iterated *implicit presence of an integrated form* throughout the operated probabilistic trials]. This significance remained hidden by the fact that, by the construction, by the rules of the game, the presence of this integrated form is constrained to manifest itself to our *knowledge* only progressively and in cryptic terms, namely via the evolving relative frequencies *{n(j)/N}, j=1,2,...q* of outcomes of this or that sign *j,* inside sequences of *N* such signs. (I make use of the word 'signs' because, by construction, *j≡Dj* and in *Dj* any trace of *form*-of-colour *D/G,κ(x_k,y_h),Vc∪𝒩(E)∪𝒩φc/, k=1,2,...10, h=1,2...10,* carried by a square, has been filtered out by the uniform 'values' of the approximate-colour aspect-view *Vac,* so that any hint of participation in a more integrated structure endowed with a global 'significance' that would exceed that description *Dj≡j,* has become non perceptible, and so in *(5)* any *Dj≡j* acts only as a sign from a set of signs). Correlatively, in consequence of the relativizations imposed by *MRC,* the vague intuitive definition of the probability of an 'outcome' acquires a quite precise character and it is associated with the *semantic* notion of 'integrated form'. Globally, the imprisonment in, exclusively, the probabilistic level of perception and conceptualization has been broken: awareness of an essential connection with another superposed level has been established.

We summarize. In the considered case, the knowledge of the colour-form carried by the global painting *P*, together with the way of parceling *P* and the view defined on the fragments, determine in *effective* terms, both, the *significance* of the assertion of *existence* of a factual probability law acting on the universe of events *{Dj}, j=1,2...q,* (namely knowledge of the presence inside the ballot box, at the time of each realization of the procedure *∏,* of the whole (parceled) painting *P*) and the *structure* of this factual probability law, namely the set *(4)* of *rational* numbers. In this case we *are* in possession of a factual interpretation-and-model for the concept of an effective factual probability law.

And by a conceptual feed-back, the interpretation-and-model obtained for this case acts as an intuitive 'justification' or 'explanation' of the theorem of large numbers. It makes this theorem immediately intelligible. It also brings into evidence the source, in *this* case, of the non effective character of the theorem, namely the impossibility to make *use* of the *finiteness* of the integrated whole (a form, in this case) that is involved in the concept of probability law *{p(e_i)} ≡ {p_F(Dj)} ≡*

---

[26] We want to strongly stress that the contents of *IV.1.5* and IV.2 are not asserted here as a *proof*, but as a *construction*. We are here outside any syntactic system. We are researching a factual definition of the concept of probability law. The theorem of large numbers has been used like a guiding element, nothing more. Our final, more general aim is precisely to clarify the *relation* between the syntactic elements and the factual elements which – together – organize the concept of probability.

*{n_P(j)/100}, i= j, j=1,2,....q*, from the theorem, simply because we do not know it and we even are *unaware* of it.

Indeed as long as one does not dispose explicitly of knowledge of *finite constraints* that determine and assign significance to the factual probability law to be asserted, while also permitting a *fragmentation in finite sequences* of the total number *N* of repetitions of the involved experiment, there is no other recourse than to make use of

*(a)* an *indefinitely* increasing integer *N* that counts uniformly, without any inner organization, the total number of achieved realizations of the involved experiment; and

*(b)* of – *directly* – a *non* effective definition of the concept of probability, in terms of a mathematical *limit* that be able to offer conceptual *room* for *any* factual sequence of *N* rational numbers.

Inside the theorem *(2)* of large numbers the assertion *(b)* of a mathematical limit for any relative frequency $n(e_j)/N$ requires *real* numbers in the formal general definition of a probability law. *Correlatively* it wholly *skips* the question of an *independent and effective* factual definition to be associated with the abstract existence of such a mathematical limit – supposing that it is genuinely useful. And furthermore it abandons inside the implicit the major fact that the formal definition, *by itself*, yields no indication whatsoever concerning the particular numerical distribution from the factual probability law to be associated with a particular factual probabilistic situation.

## IV.3. Generalization

### *IV.3.1. Preliminaries*

The 'probabilistic game' with the puzzle of the painting *P* has brought forth perception of the possibility that *any* given factual probabilistic situation be somehow connectable with a corresponding 'global form' that expresses the semantic content of the considered situation and which determines by mere counting and in terms of a finite set of *rational* numbers, the factual probability law to be asserted in that particular probabilistic situation. In what follows we investigate this possibility.

### *The role of MRC throughout VI.1 and VI.2.*

To begin with, let us note that *the new insight gained above is indelibly tied with MRC*. The development from VI.1 and IV.2 would not have been possible in the absence of an explicit awareness of the fact that *any* communicable and consensual knowledge is ***description*** and without a systematic relativization of each one among the various sorts of description that have been considered, to a well defined corresponding epistemic referential *(G,V)*, so to a definite triad *(G,œ_G,V)*. Indeed the preceding development stayed blocked a priori as long as a description *D/G,œ_G,V/* was simply ***identified*** *with the object-entity œ_G involved in it*, which furthermore was being denoted like in the set theory by just a written label *'e_i' devoid* of any semantic content specifically connected with the considered factual situation, as indeed it *is* inside Kolmogorov universes of elementary events: in such conditions no 'contours' of any sort would have been expressible, no 'local forms' nor 'global form', no 'semantic' attractions to pass from the first ones to the second ones, *nothing* available to host and to guide a puzzle-like approach, nothing available for even only being able to conceive of it. Moreover, *before* having perceived that 'truth' also is a definite concept only when it is relativized, so before having acquired capacity to confront the problems of truth in a sense that becomes *definite* in consequence of relativizations, the danger to sink into paralyzing false problems of absolute 'truth' would have threatened the mind at each step.

But now we have grown aware of the crucial role of explicit and exhaustive relativization. In particular, it has become clear that the concept of a factual probability law, since it is involved in relative descriptions like any communicable and consensual knowledge, is marked by non removable relativities. So now we know that:

> Speaking of ***'the'*** probability law p(U) to be asserted on the universe U from a probability space, is a false absolute, a huge, and devastatingly false absolute.

Therefore from now on, to avoid fuzziness and the consequences of fuzziness, we have to work on the basis of explicit general *MRC*-redefinitions. We reckon that these will clearly disclose new guiding lines – and possibly also categories of cases and *limits* – concerning the possibility to construct the unknown factual numerical probability distribution that can be asserted on a given universe of elementary events.

### *MRC definitions and consequences*

Consider what is currently called 'a random phenomenon' accordingly to Kolmogorov's *formal* theory of probabilities *(Π,U)* (cf. II.1). We want to specify the corresponding factual and relativized *MRC*-concept.

The *MRC*-definition of the procedure *Π* from a random phenomenon *(Π,U)* has been found in earlier works (MS [2002], [2006] pp. 193-202) to quite generally correspond – in *factual* terms – to a "big" number *N* of *repetitions* of the sequence of operations *[G.V]* corresponding to a given epistemic referential *(G,V)* (like in the special case of microstates). So for a factual procedure *Π* we write *Π≡[G.V]* where, for *each* realization of *Π*:

*(a) G* is a factual operation of generation which – by methodological posit – systematically re-introduces one *same* corresponding *entity-to-be-described* $œ_G$;

*(b) V* is a factual, active, operational view which re-introduces some given global 'experimental situation' and creates deliberately the whole action that leads to a qualification of $œ_G$ via aspect-values of the aspect-views *Vg* from *V* that are observable on the final effect of one complete sequence *Π≡[G.V]*. Often *V* involves a *registering test-entity*, for instance a dice[27], which is an 'object' in the usual classical sense.

By what precedes the *MRC*-redefinition of the repeatable procedure *Π* is achieved. So we can proceed to establishing the *MRC*-redefinition of the universe *U* of elementary events produced by *Π*.

In general the view *V* includes more than one aspect-view *Vg*. So 'one' realization of the sequence *Π≡[G.V]* in general involves values of several aspect-views. In the case of a microstate the aspect-views *Vg* from *V* in general are not all mutually compatible in the sense defined in III.3.2, so 'one' full realization of *Π≡[G.V]* in general consists of a union of distinct and mutually incompatible sequences *[G.Vg]*. But here, in agreement with the classical theory of probabilities, we suppose that all the aspect-views from the acting view are mutually compatible so that all the effects of the realization of one sequence *Π≡[G.V]* are obtained simultaneously. So here *[(a)+(b)]* entails that each one realization of the sequence of epistemic operations *Π≡[G.V]* produces, for the entity $œ_G$ generated by *G*, one given set of qualifications by aspect-values *gk* of the aspect-views *Vg* from *V*.

Now, according to the central *MRC*-definition *5* from III.5 the assertion of *any* relative description requires *repetitions* of the corresponding sequence *[G.V]*. (Indeed for a fully regular relativized description it can be conceived a priori that when the procedure *Π≡[G.V]* is repeated *N* times, the *same* group of *gk*-values of the aspect-views *Vg* from *V* comes out *N* times; and if that happened we would be by definition in presence of an *'N*-individual' description $D/G,œ_G,V/$ (III.5 point *5*); in this sense even an *'N*-individual' description involves *N* repetition' of *Π≡[G.V]* in order to be able to assert its *N*-individuality). But here, by hypothesis, we are *not* in presence of an 'individual' description, we are in presence of a case where it is assumed that the succession *Π≡[G.V]*, when repeated a large number *N* of times, produces a whole universe *U* of mutually distinct 'elementary events'. So in *MRC*-terms, each *one* of these elementary events can only be regarded as a sort of *limit* of the canonical *MRC*-concept of an *N*-individual relativized description, namely the limiting case in which *N=1*. Nonetheless we are now already in presence of relativized description instead of a mere notation $e_i$ devoid of any specified semantic content, as it is the case in the classical theory of probabilities; and the probabilistic game with the puzzle of a picture has offered already an opportunity for realizing how precious new guidelines for reasoning and epistemic actions are offered by the fact that inside *MRC* what is called an elementary event reemerges as a relativized description.

A limit-description of an elementary event *Dr* in the sense just specified above will be denoted $Dr/G,œ_G,V/$ or, in short, *Dr*. Since inside *MRC* everything is finite, the universe *U* of all the possible descriptions *Dr* producible with the epistemic referential *(G.V)* is finite. So the *MRC*-redefinition of the universe *U* of elementary events produced by *Π* can be written as $U≡\{Dr\}$, *r=1,2,...s*, with *s* a finite integer and *r* a *global* unique index associated to the *whole set* of qualifications *{gk}*, $\forall Vg \in V$, produced by *one* realization of the procedure *Π*.

---

[27] In general more than only one test-entity can be involved, but for simplicity here we speak of only one such entity.

The preceding considerations entail the following two *MRC*-definitions.

**MRC-definition of a factual relativized random phenomenon**. In *MRC* terms, a *factually realizable* and relativized random phenomenon *(Π,U)* has to be re-written as

$$(Π,U) \equiv \{ \ [G.V]_n, \{Dr\}, \ r=1,2,...s, \ n=1,2,....N \ \} \equiv D/G, \text{æ}_G, V/$$

where: *Π≡[G.V]*, with *G* an epistemic operation that introduces the entity-to-be-described *æ_G* ; *V=∪Vg, g=1,2,...m*, with *m* an integer, is an operational view that exists with respect to *æ_G* ; *U≡{Dr}*, *r=1,2,...s*, with *s* a finite integer and *r* a global index associated to the *whole set* of qualifications produced by *one* realization of the procedure *Π≡[G.V]*; *n* – finite – labels the repetitions of the procedure *Π≡[G.V]* and *'the'* relative description *D/G,æ_G,V/* which finally emerges (mind the singular) is a *statistical* description (III.5, point *5*) and is presupposed to be endowed with some definite though conventionally chosen *(N,N')* stability (III.5, point *5*).

**MRC-definition of a factual relativized probabilistic situation**. Consider a random phenomenon *(Π,U) ≡ {[G.V]_m {Dr}, r=1,2,...s, n=1,2,....N} ≡ D/G,æ_G, V/*. The experimental circumstances supposed by this concept – namely the 'identical' repeatability of the procedure *Π* – seem to justify the presupposition that – in a factual sense that can be shown to be in agreement with the formal non effective theorem of large numbers – the statistical relativized description *D/G,æ_G,V/* will necessarily manifest a degree of *(N,N')*-stability that permits to *posit* a 'probabilistic convergence' of the relative frequencies *n(Dr)/N* where *Dr∈U*, *r=1,2...s*, when *(N,N')* are increased. This amounts to just **posit** that in the considered circumstances a factual numerical probability distribution *{p_F(Dr)}*, *r=1,2...s* on the universe *U≡{Dr}*, *r=1,2,...s* can be identified. So we shall say that a relativized factual random phenomenon is *presupposed* to create a corresponding relativized *factual probabilistic situation*.

### Consequences

Let us now explicate the consequences of the two preceding definitions and the questions raised by these.

**(a) Descriptional aims and corresponding descriptional roles**. The preceding definitions bring into evidence the feature of freely chosen descriptional *aims* and of corresponding descriptional roles. This feature is crucially important throughout *MRC*. In the following point *(b)* it will lead us to a surprising conclusion.

In microphysics, quasi systematically, the operation of generation *G* acts inside a *basic* epistemic referential and it is *radically* creative, while the corresponding entity-to-be-described is not directly observable. But in classical probabilistic situations the operation of generation *G* is – in general – just an operation of selection, a *selector* which, each time that it is realized, re-introduces an entity-to-be-described that is *directly* observable and even can, in particular, consist of an already preexisting 'object' in the usual sense.

For instance, consider a dice-random phenomenon and let us imagine successively two different cases.

*(a1)* The dice is suspected to be loaded and the descriptional aim is to know whether yes or not this suspicion is founded. This aim suggests to form an epistemic referential *(G,V)* where *G* consists of selecting the 'object' consisting of the dice and assigning it the descriptional *role* of the entity-to-be-described, while the table on which the dice is thrown is regarded as a *registering device* incorporated to the qualifying view *V* from *(G,V)*. These choices entail that the table is supposed to be a *known* datum that has been previously worked out correspondingly to the descriptional role assigned to it (its horizontality, planarity, etc., have been insured). In this case one elementary event consists of just the number readable on the upper face of the dice when it has settled on the table after having been thrown on it. But the researched *description* of the dice is not exhausted by the individual reading of the result of only one throw of the dice, it is a statistical description. The observation of a quasi uniformity of the whole obtained statistical distribution will be coded by saying « the dice is not loaded », while non uniformity of this distribution will be coded by saying « the dice is loaded »: in this case the factual probability law on the universe of elementary events is not known and furthermore its exact form is not even researched. It simply is exterior to the chosen descriptional aim.

*(a2)* But nothing hinders to choose another descriptional aim, namely to construct a *fair* dice-game. This might induce the choice of an epistemic referential *(G,V)* where *G* is a selector which now selects the table to be put in the role of the entity-to-be-described (in order to check its degree of horizontality and of smoothness) while this time the dice is assigned the role of a registering device incorporated to the acting view. So this time the geometry and structure of the dice have been deliberately constructed before such as not to favor by its inner structure any one among the six possible outcomes *1, 2...6*. But we do not know whether the table also is adequate, so for this we have to check. In these new conditions an elementary event consists of the place on the table where the dice settles after a throw. And the researched description of this place is statistical again: uniformity of the obtained statistical distribution of the settled locations of the dice on the table will mean « the table is plane and smooth » while non-uniformity will mean « the table has defaults of smoothness or/and planarity ». In this case the factual probability distribution on the universe of elementary events is neither known, nor unknown: according to the chosen descriptional aim this law has to be *constructed* such as to be uniformly distributed. (Obviously the roles of the dice and the table can be reversed, the aim of constructing a fair dice-game being kept invariant).

The considerations from *(a1)* and *(a2)* suffice for illustrating in what a sense according to *MRC* 'operation of generation', 'object-entity-to-be-studied', 'view', 'elementary event', 'description', are not features tied with some support, but only *descriptional roles* corresponding to the choice of an epistemic referential (cf. III, point *7*).

This sort of *in*dependence of any fixed support – whether material or conceptual – that characterizes the *MRC* descriptional roles *"G", "œ_G", "V"*, founds the flexibility of *MRC*, the possibility to apply it for any descriptional aim (MMS [2006] pp. 97-105). It can be regarded as a sort of methodological 'principle of relativity' that insures an *invariant* general procedure for describing in *MRC*-terms whatever is describable and one wants to describe.

*(b) The concept of random phenomenon as a methodological artefact*. The *MRC*-definition of a factual and relativized random phenomenon with the corresponding 'probabilistic situation', together wit the comments from *(a)* bring into evidence a fact that might come as a surprise, though as soon as it has been formulated it appears as obvious:

A formal 'random phenomenon' is *not* a concept that designates a *naturally* existing structure. It is a general *conceptual-factual artefact*. And each specified factual random phenomenon – in particular, as defined inside *MRC* – is a conceptual-factual artefact deliberately constructed with a corresponding *aim*; namely precisely the *creation* of a corresponding 'probabilistic' factual situation, that is, a situation endowed with 'probabilistic predictability', so involving a definite factual numerical distribution of probabilities coherent with the general formal concept of a probability law.

One might be tempted to resist this assertion. One might want to think that factual random phenomena with their respective factual probabilistic situations can also occur naturally. This doubt, however, is doomed to fade away. One realizes soon that a full and rigorous materialization of the concept of random phenomenon cannot be conceived as a natural circumstance. Natural conditions can only yield a *basis* for organizing a random phenomenon out of naturally statistical impacts upon the human biological registering apparatuses, with their artificial prolongations.

Consider for instance the case of meteorological predictions. The earth with its water volumes-and-surfaces, its relief, vegetation, fauna, and its atmosphere, constitute a finite and practically closed whole which nowadays has become perceptible as such by the help of global descriptions realized from satellites. Let us consider a relativized description $D(t_1)$ of this whole achieved at a time $t_1$. The view from $D(t_1)$ is a 'meteorological view' including the aspect views of "pressure", "temperature", "humidity", "local velocity of displacement", etc., as well the space time frame-aspects $E$ and $T$ (the value $t_1$ of $T$ being fixed). The Navier-Stokes equations permit to transpose $D(t_1)$ into a set of other descriptions $D(t_2)$, $D(t_3)$... $D(t_q)$... corresponding to a set of other time-values $t_1 > t_2 > t_3 .... > t_q ,...$, the utilized view remaining fixed. If these time-values are all sufficiently close to $t_1$ the transpositions are endowed with a remarkable degree of certainty. But when the time interval $t_q$-$t_1$ is increased this degree of certainty decreases and one is pushed toward probabilistic predictions.

Now, a human being does never *directly* perceive a global description from the set $D(t_1)$, $D(t_2)$, $D(t_3)$... $D(t_q)$.... At any given time $t_q$ he can only encompass a spatially local perception of the corresponding global description $D(t_q)$. Though in this case $D(t_q)$ *itself* is not parceled like in the case of the picture $P$, the description $D(t_q)$ can be treated as an equivalent of the global description of $P$ from IV.1. Indeed, suppose that on the basis of the knowledge of only $D(t_1)$ we want to make a prediction concerning the various local perceptions accessible to a human being at a time $t_q$, the unit of time being, say, one day. Then $t_q$ is relatively distant from $t_1$ and in the present state of our meteorological knowledge the prediction cannot be certain. But a probabilistic prediction can be researched. In particular it can be researched via a "probabilization" of $D(t_q)$. The global description $D(t_q)$ (just like that of the picture $P$) can be divided in $n_D$ fragments (for instance it can be realized such that each fragment covers an area equal to that of town of medium extension). Each one of these fragments emerges as a relative description $Dr(t_q)$ achieved with respect to the same aspect-views as $D(t_1)$ but which is local with respect to the frame-aspect $E$ of space, $r$ being a unique notation for the whole group of values of the aspect-views from the meteorological view from $D(t_1)$ that is realized on the considered local description. In these conditions one can work with the descriptions $D(t_q)$ and $Dr(t_q)$, $r=1,2...s$ in a way entirely similar to that from our example of the puzzle of the painting $P$.

So one might think that in this case, via the procedure from VI.1, it is possible to identify the probability distribution on the local descriptions $Dr(t_q)$, $r=1,2...s$ by working with the *naturally* preexisting initial situation denoted $D(t_1)$ and with the subsequent *natural* statistical local manifestations denoted $Dr(t_q)$, $r=1,2...s$. But obviously the jump to such a conclusion cannot be accepted. For the treatment outlined above involves quite essentially relativized *descriptions* (among which the Navier-Stokes equation which also is description) and descriptions are not natural facts, they have to be deliberately realized and to be manipulated in a way *such* as to permit to make probabilistic predictions on weather.

So we reassert that the general *MRC* concept of a factual random phenomenon is always an artefact: it is a **methodological artefact** conceived for capturing a sample from the ocean of natural randomness where we are immersed and for subjecting this sample to local constraints which – in specific relation with this or that particular aim – do insure the more or less feeble – but not null – degree of predictability, endowed with a peculiar sort of stability, which is called probabilistic predictability.

This conclusion is a guide, in various ways. It brings into evidence dimensions of liberty for *deliberate* conceptual-factual constructions which can be fully exploited if – and only if – they are clearly known as such; whereas if they get mixed with obscure tendencies to 'discover' some 'independently preexisting factual truth' they can only produce indefinite stagnation and even misleading assertions and procedures.

For instance, let *(Π,U)* be a factual random phenomenon involved by a "fair" game of chance. It is deliberately constructed as such under specific constraints. Among these let us note the fact that in this case *the randomness is deliberately included in the procedure Π*, in a way which (in general at least) insures – for the **own** features of randomness – a 'normal' Gaussian statistical distributions, without thereby altering the distribution imposed upon the factual probabilities on the elementary outcomes. This draws attention upon a frequent *confusion* between:

- The distribution of numerical probabilities assigned to (or constructed for) the random features *of the modalities of performing the experiment Π.*

- The distribution of numerical probabilities to be asserted on the universe $U$ of the events produced by the repetitions of the experiment *Π.*

*This sort of confusion might explain the unreasonable profusion of Gaussian distributions asserted on the universe U of the events produced by the repetitions of the experiment Π* (It is Jean-Marie Fessler [2008] who, in a private exchange, has strongly drawn my attention upon this unreasonable and certainly false profusion of Gaussian distributions asserted on universes $U$ to be studied).

**(d) On the 'significance' of the 'existence' of a factual probability distribution**. But what sort of significance, exactly, can be assigned to the *mere* 'existence' of the factual probability law entailed by a given probabilistic situation, abstraction being made of its form?

Karl Popper's well known 'propensity interpretation' (MS [1992B], [2002], [2006]) offers a first rather strong indication:

"Take for example an ordinary symmetrical pin board, so constructed that if we let a number of little balls roll down, they will (ideally) form a normal distribution curve. This curve will represent the *probability distribution* for each single experiment, with each single ball, of reaching a possible resting place. Now let us "kick" this board; say, by slightly lifting its left side. Then we also kick the propensity, and the probability distribution,... Or let us, instead, remove *one pin*. This will alter the probability for every single experiment with every single ball, *whether or not the ball actually comes near the place from which we removed the pin.* .....we may ask: "How can the ball 'know' that a pin has been removed if it never comes near the place?" The answer is: the ball does not "know"; but the board as a whole "knows", and changes the probability distribution, or the *propensity*, for *every* ball; a fact that can be tested by statistical tests".

According to this interpretation the global experimental situation introduced by the procedure *Π* from any given random phenomenon, with all the material objects and all the actions involved by it, determines the specifically corresponding numerical law of distribution of the probabilities of the involved elementary events.

We admit this interpretation as fully general. But in what follows we shall try to bring into evidence the whole structure of its foundation; and then, on this basis, to construct a general *effective* procedure for specifying the factual numerical distribution of probabilities involved by a probabilistic situation, in coherence with the formal theorem of large numbers. Finally, we shall submit to discussion the question of the limits of this effective procedure, so the question of the limits to be assigned to the constructability of a 'probabilistic situation' in the full sense of this expression.

### IV.3.2. Probabilization of any given entity perceived or conceived as a whole

We begin with the easiest part, namely the generalization of the procedure which in IV.1 led from the picture *P* to a corresponding probability game.

According to *MRC* «…only descriptions are *unrestrictedly* communicable **knowledge**, not 'facts' which are exterior to any psyche, nor psychic facts (emotions, desires, etc.) which are not expressed by some more or less communicational structure, verbal or of some other constitution. When we say « I know this house » we spell out an illusion, mainly because of unawareness but also for the sake of brevity. Indeed only the assertion « I know the *descriptions* (plural) of this house » would correctly express the situation toward which we wanted to point (III.5, *MRC*, point *(11)*). Furthermore, any relative description $D/G,æ_G,V/$ consists of a cloudy structure or 'form' *of points* from the *m*-dimensional representation-space of the view *V* from $D/G,æ_G,V/$, points representing *(gk)*-values with *g=1,2...m* and *k=1,2,...w(g)*. If in particular the entity-to-be-described $æ_G$ is of a physical nature one must add to *V* a space-time view *V(ET)* (III.5, point *6*). Obviously any such global whole or 'form' of points, if it is introduced as the *primary* datum, *can* be subject to *various* modalities of fragmentation, so to various relativized 'probabilizations', each one realized in ways quite similar in essence to that performed in IV.1. So the procedure of probabilization *can* be generalization to any relative description $D/G,æ_G,V/$ that is introduced as a primary datum.

This possibility, though in many circumstances it is of no use, might play a quite important role in certain domains (like medicine (systematic relativization of the scanning procedures and other modalities of analytic investigation, military explorations, explorations conducted from satellites or spatial devices or by robots, etc).

### IV.3.3. Toward a general procedure for constructing factual probability distributions

Conversely now, we *start* with the relativized *MRC*-description of a factual probabilistic situation generated by a given random phenomenon and from the data offered by this description we shall try to specify the corresponding factual numerical distribution of probabilities via an effective procedure similar to that from IV.1 and IV.2. With respect to this new descriptional aim the path to follow is much less obvious. So we develop a very progressive approach.

*Distillation of the essential points*

*Relatively 'simplified' and relatively 'complexified'.* Let us once more go back to the example of the probability game with the puzzle of the painting *P*. The aim is to bring into evidence which features of the approach practiced in that particular case are essential for a generalization and which ones can be dropped.

The universe of *label*-elementary-event-descriptions $U \equiv Dj$, $j=1,2,....q$ from that example consisted of a set of *q* mutually distinct sorts of pieces of puzzle – with $q \ll 100$ – that had been *extracted* (in certain specified conditions) from the set of *100* 'local' relative descriptions $D/G_\sigma, \sigma, V(El) \smile c\phi/$ constituting the pieces of the puzzle of *P*. These 'local' descriptions $D/G_\sigma, \sigma, V(El) \smile c\phi/$ were maximally individualized with respect to the descriptional potentialities of the epistemic referential $(G_\sigma, V(El) \smile c\phi)$ where they had been constructed. Whereas the set of descriptions *{Dj}, j=1,2....q* had been obtained by re-qualifying each one of all these 'local', maximally individualized relative descriptions $D/G_\sigma, \sigma, V(El) \smile c\phi/$, by the use of a *simplifying view Vac* of uniform approximate-colour. This, because any *MRC*-view is by definition a *filter* (III.5, point *4*), has reduced the *perception* of each local relative description $D/G_\sigma, \sigma, V(El) \smile c\phi/$, to only a unique approximate-colour value *j* uniformly spread over it. So when, instead of making use of the 'complete' view $V(El) \smile c\phi$ acting inside the referential $(G_\sigma, V(El) \smile c\phi)$, we decided to make use for *looking* at a local description $D/G_\sigma, \sigma, V(El) \smile c\phi/$, of only the simplifying view *Vac*, we were led to just re-write this 'local' description as a simplified 'label-description' $D/G_\sigma, \sigma, Vac/ \equiv Dj$.

On the new simplified universe $U \equiv \{D/G_\sigma, \sigma, Vac/\} \equiv \{Dj\}$, $j=1,2,....q$ where we had $q \ll 100$, any connection with the integrated relative description $D/G_{P_t}, P, V(El) \smile c\phi/$ of the whole painting *P* had been *effaced*, it had been replaced by a *cut* from this integrated description of *P*.

*Correlatively* inside any probability space $[\{Dj\}, \tau_x, \{p(Dj)\}]$, $j=1,2,....q$ constructed on $U \equiv \{Dj\}$ the factual numerical probability distribution *{pF(Dj)}, j=1,2,....q* to be inserted into the *abstract* measure *{p(Dj)}, j=1,2,....q*, was *unknown*. But in that particular case, the theorem of large numbers associated with the *knowledge* of the integrated puzzle of the painting *P*, led to the assertion of the researched numerical factual probability distribution $\{p_F(Dj)\} \equiv \{n_F(j)/n_{PT}\} \equiv \{n_F(j)/100\}$, $j=1,2,....q$. This happened in consequence of the very particular circumstance that each label-elementary-event-description *Dj* had remained *materially* immersed in an *already available* and previously perceived more complex 'local' relativized description $D/G_\sigma, \sigma, V(El) \smile c\phi/$ (the operation of generation $G_\sigma$ remained unchanged) that was itself a *material fragment* of the integrated description $D/G_P, P, V(El) \smile c\phi/$ of the whole painting *P*. This compensated for the cut mentioned above, in the following way: it permitted to *refer* the 'simplified' descriptions from $U \equiv \{Dj\}$ to the more complex local relativized descriptions $D/G_\sigma, \sigma, V(El) \smile c\phi/$ from the integrated puzzle of *P* and to then to count the occurrences of each *Dj* in a description $D/G_\sigma, \sigma, V(El) \smile c\phi/$ from the puzzle of *P*. And this in its turn permitted to specify the factual probability distribution $\{p_F(Dj)\} \equiv \{n_F(j)/n_{PT}\} \equiv \{n_F(j)/100\}$, $j=1,2,....q$ by comparison with the theorem of large numbers.

But it seems certain that so propitious conditions do not realize in general. So, could one compensate *in general* for the absence of perceivable connections with a possibly still not materially constituted global form permitting to determine in an effective way the researched factual numerical distribution of probabilities? *What, exactly, in the case of the puzzle of P, did these connections entail, which was so essential for revealing the researched factual distribution of probabilities, and what can be dropped?*

Let us first focus our attention upon the set *{Dj}, j=1,2,....q*, of 'label-descriptions', with $q \ll 100$. This amounts to starting this time with a random phenomenon that produces directly the universe $U \equiv \{Dj\}$, $j=1,2,....q$). *With respect* to this set *{Dj}, j=1,2,....q*, the set of *100* 'local' descriptions $D/G_\sigma, \sigma, V(El) \smile c\phi/$ worked out inside the referential $(G_\sigma, V(El) \smile c\phi)$ is a 'complexified' set; a set which is richer by both the semantic content assigned to each element and by the number of elements. From a probabilistic point of view, this richer set of the descriptions $D/G_\sigma,\sigma,V(El)\cup c\phi/$ can be regarded as a 'universe' $U^\complement$ of *100* complexified 'elementary events' on which one can construct the complexified factual probability *space $[U^\complement,\acute{\tau}_T,\ p_F(\acute{\tau}_T)]$* where the algebra is chosen to be the **total** algebra on $U^\complement$ so that we have $U^\complement \subset \acute{\tau}_T$ and $p_F(\acute{\tau}_T)$ is the factual numerical distribution of probabilities *directly defined on* $\acute{\tau}_T$ (as it is required by the abstract concept of a probability space) but which contains also the probability measure on $U^\complement$. Now, the complexified probability space $[U^\complement,\acute{\tau}_T,p_F(\acute{\tau}_T)]$ possesses a feature which in the present context is important:

> Any description $Dj\,\epsilon\,U$ which inside the factual probability space $[U\equiv\{Dj\},\tau_X,\ p(\tau_X)]$, $j=1,2,....q$ constructed on $U$ was an *elementary*-event-description, reappears inside $\acute{\tau}_T$ as an *event*-description that is 'realized' by a whole set of complexified elementary-event-descriptions $D/G_\sigma,\sigma,V(El)\cup c\phi/$ from the universe $U^\complement$.

*MRC* brings into strong evidence the well known but little stressed *relativity* of the elementary character of an event, in the probabilistic sense: elementary character, or not, of a probabilistic event is only a matter of *reference*. This entails a consequence. Inside the probability space $[U^\complement,\acute{\tau}_T,p(\acute{\tau}_T)]$ which, with respect to the initially considered space $[U\equiv\{Dj\},\tau,\ p(\tau)]$, $j=1,2,....q$ is 'complexified', one can *count* the number of complexified elementary-event-descriptions $D/G_\sigma,\sigma,V(El)\cup c\phi/$ from $U^\complement$ that realize a given considered description $Dj$ from the complexified total algebra $\acute{\tau}_T$ constructed on $U^\complement$ (which are 'favourable' to the outcome described by $Dj$ because inside $[[U^\complement,\acute{\tau}_T,p(\acute{\tau}_T)]]$ an outcome $Dj$ is less constraint than an outcome $D/G_\sigma,\ \sigma,\ V(El)\cup c\phi/$). It was on the basis of *this* fact that, inside the integrated puzzle of the picture $P$ which represents precisely the 'complexified' universe $U^\complement$, we have been able to identify the ratios $\{n_P(j)/n_{PT}\}$, $j=1,2,...q$ (which then have determined the effective numerical factual distribution of probabilities $\{p_F(Dj)\}\equiv\{n_P(j)/n_{PT}\}\equiv\{n_P\ (j)/100\}$, $j=1,2,....q$, *(F: factual)* from *(4)* by mere identification of terms with the equation (2) expressing the theorem of large numbers).

> The generation of ***a reference*-background $U^\complement$ endowed with a structure specified with more detail than that of the initially considered descriptions $Dj$ from $U$ themselves**, permitted specifications concerning the descriptions $Dj$ which in the absence of such a reference-background were not possible.

Let us now note that we would have obtained the same result inside the classical theory of probabilities – without making use of the integrated puzzle of P – if we had directly posited that inside the more complex space $[[U^\complement,\acute{\tau}_T,p(\acute{\tau}_T)]]$ the probability of an event $Dj$ from $\acute{\tau}_T$ was *proportional* to the number of mutually distinct complexified elementary event $D/G_\sigma,\sigma,V(El)\cup c\phi/$ from $U^\complement$ that realized $Dj$. But this posit would have been *equivalent* to postulating that the distribution of probabilities on the universe $U^\complement$ was *uniform*. Indeed this postulation would have entailed precisely the numerical factual distribution of probabilities from *(4)*, namely – on the basis of the general formal requirement $p(A\cup B)\leq p(A)+p(B)$ where *equality obtains* iff $A$ and $B$ are mutually 'independent' (cf. note 3):

> This postulation would have rendered *superfluous* the possibility to act with the fragmentary local descriptions $D/G_\sigma,\sigma,V(El)\cup c\phi/$ from $U^\complement$ as in a puzzle game, so also the existence of 'semantic continuities' on the borders of these.

But why should we postulate a uniform factual distribution of probabilities on the universe $U^\complement$?

***[a priori - a posteriori] oscillations in classical probabilities.*** Inside the domain of the classical factual probabilistic conceptualization, Laplace has introduced his well known principle which requires the a priori assignation of a uniform distribution of probabilities on the universe $U$ of elementary events from any space $[U,\tau_X,\ p(\tau_X)]$. This a priori assignation, however, is in Laplace's thinking just an *initial bet*. This bet is expected to be in general invalidated by a posteriori measurements of the relative frequencies of the considered elementary outcomes from $U$; and when this happens indeed, and often it does, it requires a re-definition of the universe $U$ of elementary events itself as well as a new bet of uniformity concerning the probabilities on this modified universe, and then a new confrontation with factuality, and so on and on. The number of necessary such pairs [a priori assumption – a posteriori confrontation] is not predictable. Sometimes this sort of [a priori-a posteriori] confrontation is repeated until the initially found dissention is effaced. But often an a priori posited uniform law simply is accepted in a definitive way, unquestioned and in absence of any a posteriori control[28] (like in Boltzmann's statistical theory of gases and presumably in many quantum mechanical investigations where symmetries are invoked). The process might even never find an end. The procedure is practiced in an arty-crafty way.

The abstract theory of probabilities has left place for the principle of Laplace. Indeed the fact that in an abstract probability space *[U, $\tau_X$, p( $\tau_X$)]* the abstract probability law $p(\tau_X)$ is defined directly on the *algebra* of events $\tau_X$, *not* on the universe of elementary events $U$ (cf. II.1), indicates that the distribution of probabilities on $U$ is considered to be always decided a priori. (This also explains the fact noticed in VI.2 that the theorem of large numbers is compatible with the intuitive definition of the probability of an 'outcome' as [the number of 'favorable cases']/[the number of all the possible 'cases'] which amounts to assuming not only $p(A \cup B) \leq p(A) + p(B)$) but also a uniform distribution on $U$.

So *the problem of the identification of a true factual numerical distribution of probabilities on U is left open by the principle of Laplace*. (Similar assertions hold concerning the well known principle of Jaynes).

*[A priori - a posteriori] oscillations inside MRC.* In the present approach however we have placed ourselves inside *MRC*, not inside the classical theory of probabilities. Does this not change the conceptual situation? Well, no, not with respect to the necessity to assume uniformity of the numerical probability distribution on a universe of elementary events: *According to MRC the principle of Laplace is a deductive consequence of the principle of separation PS*[29], and this last principle plays a quite crucial role throughout *MRC*.

---

[28] When this is done *the 'predictions' drawn from the calculi are in fact only 'best bets'.* And very often this is the case indeed, for even if the involved 'constraints' (in particular the 'limiting conditions') *are* taken into account and so the principle of Jaynes is invoked instead of the principle of Laplace, still what is accepted is a best a priori bet that remains open to a posteriori invalidation.

[29] Inside *MRC* – quite remarkably – the principle of Laplace follows *deductively* from the principle of separation *PS*:

**Proposition**. Consider a random phenomenon *(П,U)*, $U \equiv [Dr]$, *r=1,2,....s* worked out inside an epistemic referential *(G,V)* where the view *V* contains *all* the aspect-views that are *directly observable* on the outcomes of *(П,U)* and the descriptions *Dr* are *maximally* worked out with respect to *V*. Consider also the probability space *[U≡[Dr], $\tau_X$(U), p($\tau_X$)]* founded on *(П,U)* where $\tau_X$(U) is the *total* algebra on *U*, so it contains also all the elementary events from *U*. Consider now the factual numerical probability distribution *p(U)* which, by construction, is contained in *p( $\tau_X$)*. In the above conditions – a priori – *p(U)* can only be the uniform distribution *p(U)≡[p(Dr)]≡[1/s], s times]*, where *s* denotes the cardinal of the index-set *r=1,2,....s*.

**Proof**. By hypothesis the descriptions *Dr* from the universe *U≡[Dr]*, *r=1,2,....s* are maximally qualified with respect to the view *V* from the involved epistemic referential *(G,V)*. So once the relative descriptions *Dr* have all been fully worked out, all the qualifying powers of *(G,V)* have been **exhausted**. Suppose now that one wants to furthermore meta-qualify these already achieved descriptions by *global* numerical qualifications, namely in the *statistical* terms of the relative frequencies of outcomes of the elementary-event-descriptions *Dr* from *U*, and then furthermore in also meta-meta *probabilistic* terms of convergence of these relative frequencies. Then one has to construct successively two new epistemic referentials, namely: first *(G',V')* where *G'≠G* introduces as [object-entity-of-description $\alpha_{G'}$-] the whole previously achieved description of *U≡[Dr]* considered globally and *V''≠V* is now a purely 'counting' statistical view that specifies values *n(Dr)/N* the aspect *g≡'relative frequency'* (with *N* the total number essays) but leaves *unchanged* the semantic contents of the descriptions *Dr* from *U* achieved inside *(G,V)*; and then a referential *(G'',V'')* where *G''* introduces as [object-entity-of-description $\alpha_{G''}$-] the whole previously achieved descriptions of relative frequencies while the view *V''* qualifies the *convergence* of these but leaves unchanged the semantic contents of the descriptions previously achieved inside *(G,V)* and *(G',V')*.

Now, according to the principle of separation *PS*, each one of these new desired descriptions has to be carried out inside its *own* epistemic referential in a way *strictly SEPARATED from the descriptional processes from the other referentials).*

In these conditions, the a priori assertion of a factual probability law *[p(Dr)], r=1,2,....s* on *U≡[Dr], r=1,2,....s* that would *not* be uniformly distributed, could *only* stem from :

(a) *suppression* – in the view *V* from *(G,V)* – of one or more aspect-views *Vg* or of one or more values *gk* of aspect-view ;

(b) *addition* – in the view *V* from *(G,V)* – of one or more aspect-views *Vg* or of one or more values *gk* of aspect-view.

But both *(a)* or *(b)* would amount to a surreptitious retroactive modification introduced in the epistemic referential *(G,V)* on the basis of some incentive induced by some other description from another referential. This would violate the principle of separation *PS*.

So the principle of separation entails with *necessity* the a priori assertion of a uniformly distributed factual probability law *p(U)≡[p(Dr)]≡[1/s, s times]*, thereby insuring that the amount of data available inside the initial epistemic referential *(G,V)*, is not arbitrarily transgressed. ∎

Now, given that the principle of separation involves subjective epistemological-methodological features, the *MRC*-requirement of uniformity of the probability distribution posited on a universe of elementary events cannot be hoped to certainly entail a posteriori confirmation of its *factual truth*. This entails that the a posteriori factual truth of the distribution of probabilities on the events from the algebra of events $\tau_X$ constructed on *U* is equally left open. In short, inside *MRC* also, as long as we rely on the methodologically required a priori assumptions of uniform distributions of numerical probabilities, we do not get rid of possibly indefinite oscillations. But this sort of oscillations are a form of non-effectiveness, precisely what we want to eliminate.

*Consequence.* The above considerations entail that in order to identify in an effective way the factually true numerical distribution of probabilities on the universe of events introduced by a given factual random phenomenon we must *circumvent any insertion of a priori assumptions: only a purely factual and finite sequence of operations could realize this aim.*

Now, in the case of the puzzle of the picture *P* effectiveness was entailed by the possibility of *counting* elementary events and events inside any replica of the *finite, closed whole 'P'*. The semantic attractions on the borders of the 'complex' and 'local' descriptions $D/G_\sigma,\sigma,V(El)\cup Nc\phi/$ have been useful for that only in so far that:
- They have permitted to build the finite closed whole *'P'* ;
- They have permitted to *know* when a replica of this whole had been completed, thus breaking retroactively the non interrupted flux of the increasing number *N* of registered 'outcomes' involved by the theorem of large numbers, into a sequence of replicas of the integrated puzzle of *P*; these were emerging in a mixed way, but separately from one another, and each one of the replicas – as soon as it had been completed – *sufficed* for estimating inside it the numbers which determined the researched probability distribution.

*Any* substitute to a puzzle game which would introduce the two features explicated above, would bring solution to Kolmogorov's aporia: *The requirement of possibility of a fully regular puzzle game, can be* **dropped** *in favor of such a minimal substitute, necessary and sufficient for our aim*. This is a liberating conclusion.

*On 'syntax' in the usual sense, factuality, and MRC.* We add now the following general remarks. We have already much stressed that inside the mathematical theory of probabilities – a *purely* syntactic system – the concept of a probability law is defined only by its general features. No statement is made concerning *what* probability has to be assigned to this or that elementary event or event from a given factual probability space. The formal definition remains void of numerical specifications. This is only a particular manifestation of a quite general and much too neglected characteristic of any formal syntax, mathematical or only logical, that is made use of in order to represent rigorously a domain of facts: The factual data to be introduced into the syntactic framework offered by such a purely formal system, *always* have to be **entirely** obtained inside the considered domain of facts.

Nevertheless there persists a widespread and strong tendency to want to *derive* factual data inside a mathematical syntax. When such a tendency is yielded to, it generates paradoxes and false problems, so stagnation. *MRC* expressly eradicates by construction the possibility of realization of such erratic tendencies, at various crucial levels (III.5 points *5* and *10*). This eradication imposes conditions of possibility: Inside any formal representation it requires 'hosting places' offered for introducing in them the necessary specifications which determine this or that factual application.

Consider now the case of Kolmogorov's classical formal representation of the concept of probability. In this representation the void formal 'hosting places' offered for introducing in them the factual elementary events or events, are far from being endowed with a structure sufficiently definite and detailed for permitting to satisfy the *MRC*-requirements. Indeed since any knowledge that is communicable without restriction, is *description*, an elementary event or an event has to be specified as a description, namely a relativized description (cf.III.5, point 12). But a relativized description is referred to a triad *(G,α$_G$,V)* and any one aspect-view *Vg* from the involved view *V* introduces *two* indexes (formal places), one for the involved semantic dimension *g* and a second one for each

---

corresponding value index *(gk)*. So, obviously, the bare notations $e_i$ and $e$ simply cannot be utilized without savagely amputating the communicable expression of the involved descriptions. These – *which have to be generated in a purely factual way* – do not find a pre-organized formal recipient where to be inserted. This withstands the full realization of applications of the classical theory of probabilities. The example with the puzzle of the picture *P* indicates how *MRC* fills in this gap.

### IV.3.4. The algorithm of semantic integration of the factual distribution of probabilities in a given probabilistic situation

The above preliminary considerations have distilled the essential features on which it can be tried to found a general procedure for estimating factual probabilities. With these features in mind we want to construct now a general for identifying the numerical probability distribution to be asserted on a factual probabilistic situation. We require this procedure to contain *nothing else* than directly a large number of factual repetitions of the procedure random phenomenon *(Π,U), U≡{Dr}, r=1,2,....s*: No a priori assumptions nor any sort of [a priori-a posteriori] dialogue; from the start and up to the end only registration of factual data endowed by construction with features permitting to finally perform the researched identification exclusively by finite conting of factually registered outcomes.

For only the first constructive step, namely the generation of a minimal substitute for the puzzle game of the picture *P* – we begin by an example. Then we shall construct the whole procedure in fully general terms.

**Introductory example of a minimal substitute for the puzzle game with P.** Let *(Π,U)_D* (*D*: dice) denote a dice-random-phenomenon constructed inside an epistemic referential denoted *(G,Vr)* where:

- *G* introduces one replica of the object-entity-of-description *œ_G* consisting of a *dice*. This dice, however is here suspected of being *loaded* and so it is treated as an *unknown* object-entity to be studied and qualified: the dice-random phenomenon considered here, instead of belonging to the domain of games, is turned into a procedure for achieving a piece of scientific research.

- *Vr* is an 'active' view that consists of performing the realizations of the experiment *Π* from the random phenomenon *(Π,U)_D* in deliberately constructed conditions, namely: The table is put in the role of a device incorporated to *Vr* and the smoothness, horizontality, etc. of the tabletop have been thoroughly controlled in advance; *Vr* includes also an automatic device for throwing the dice; each time that, after a throw, the dice has settled still on the table, the numerical value of a global index *r=1,2,3,4,5,6* is by definition the number marked on the upper face of the dice; a very big number of previous tests has established that the dimensions of the tabletop and the mechanism for automatic throws are *such* that the set of all the possible locations where the dice can settle after a throw is confined inside a finite *spatial* zone on the table-top, which is largely distant from the edges. Let us denote this zone by *Z*.

The universe generated by *(Π,U)_D* is by definition the set *U≡{Dr}, r=1,2,3,4,5,6* of the specifications of a value of *r*. On this universe we construct the probability space *[U≡{Dr},τ_f(U),p_F(U)], r=1,2,3,4,5,6* where *τ_f(U)* is the *total* algebra on *U* and the unknown factual distribution of probabilities which we want to identify is *p_F(U)≡{p_F(Dr)}, r=1,2,3,4,5,6*. For simplicity *{p_F(Dr)}, r=1,2,3,4,5,6* is researched directly on *U* (which then determines it also on *τ_f(U)*).

Let us now construct for the universe *U* to be studied, a 'complexified' structure of embedding and reference. Let *(G,V^c)* (*c*: complexifed, complexifying) be a epistemic referential where the operation of generation *G* is the same as in *(G,Vr)* but the complexified view *V^c* perceives *all* that in this case is *directly observable* concerning the outcome of a dice-throw, not only the six label-numbers *r=1,2,3,4,5,6*, namely:

- The values of the label-aspect-view *Vr*.

- The projection on the tabletop of the position of the centre of gravity of the dice with respect to a definite frame-aspect-view of space *VE*.

- The orientation of the dice with respect to the directions of the edges of the rectangular table; this can be defined, say, by the angle *α* between a reference direction parallel to a given edge of the table, and an edge of a face of the dice singularized by some explicit definition.

So we have now $V^c \equiv Vr \cup N \alpha \cup NE$. The new complexified epistemic referential $(G, V^c)$ with $V^c \equiv Vr \cup N \alpha \cup NE$ generates a complexified set of elementary-event-descriptions, $U^c \equiv \{Dr^c\}$, $r^c = 1,2,....s^c$. Oviously $s^c > s$. Now – and this is quite essential in the present context – nothing permits to assume that all the complexified representations $Dr^c$ of this or that initial description $Dr$ do factually occur when the initial experiment $\Pi$ is repeated an arbitrarily large number of times: We are not informed of the physical determinations which entail factual possibility or factual exclusion of a given complexified outcome $Dr^c$ such as it is singularized by the complexified view $V^c$. The random phenomenon $(\Pi, U)$ has been constructed under deliberate controls which did not take into account all the aspects and values of aspects subsequently introduced in the complexified view $V^c$. These controls are quite likely to have entailed physical characters involved in $\Pi$ of which the roughness or even the nature excludes the factual outcome of certain $Dr^c$ even though these are *conceptually* constructible with respect to the complexified view $V^c$; while among the complexifications $Dr^c$ which have remained factually realizable, the outcomes are factually favored or inhibited by the mentioned physical characters in different unknown ways. In short, *the **factual** content of the complexified $U^c \equiv \{Dr^c\}$, $r^c = 1,2,....s^c$ and the factual distribution of probabilities on it, are not known in advance*.

Let us now achieve a more intuitive perception of the whole conceptual content of $U^c$. This can be done best by constructing an abstract representation space of $V^c \equiv Vr \cup N \alpha \cup NE$. We choose to subtend the complexified view $V^c$ by mutually orthogonal semantic dimensions. Then the spatial location frame-aspect-view $VE$ introduces three mutually orthogonal semantic reference-directions, $ox$, $oy$, $oz$, that carry on them, respectively, values $x$, $y$, $z$, defined with some finite arbitrary precision, say corresponding to a unit of length of $5$ millimeters. The label aspect-view $Vr$ introduces one semantic label-direction that carries displayed on it the six 'values' $1,2,3,4,5,6$ (marked in an arbitrary order). The orientation aspect-view $V\alpha$ introduces a fifth semantic direction that carries on it values of the angle $\alpha$ equally defined with some finite arbitrary precision, say corresponding to a unit of $5$ degrees.

According to *MRC* (III.5, point *(4)*) any aspect-view is a filter that is blind with respect to qualifications that are not built into it. So the view $V^c \equiv Vr \cup N \alpha \cup NE$, endowed with the chosen *units* of length and of angle, does not distinguish qualifications that lie inside a unit interval. Let us agree that such a qualification is assigned the value defined by the biggest one of the two numbers that delimit the involved unit interval. This leads finally to a well defined $5$-dimensional representation space endowed with 5 mutually orthogonal semantic directions and defining a finite *points*-grid of valuation. Let $\gamma(V^c)$ denote this points-grid. Then a *given* quintuple of values of the parameters $r,x,y,z,\alpha$ defines one given 5-dimensional point from $\gamma(V^c)$ represented by the *global* value-index $r^c(r,x,y,z,\alpha)$. So each *conceptually* constructed complexified elementary-event-description $Dr^c \in U^c$ is represented by a point $r^c(r,x,y,z,\alpha)$ from the grid $\gamma(V^c)$, and vice versa.

We now construct the total algebra $\tau^c_T$ on $U^c$. The *elementary*-event-descriptions $Dr \in U$ from the initial universe $U \equiv \{Dr\}$, $r = 1,2,3,4,5,6$ have now migrated into $\tau^c_T$. And there – with respect to the complexified view $V^c \equiv Vr \cup N \alpha \cup NE$ – they appear retroactively as 'simplified' *event*-descriptions reduced to exclusively qualifications via the aspect-view $Vr \in V^c$. So now, with respect to $V^c$, each given one outcome $Dr \in U$ appears as realizable via a whole set of mutually distinct descriptions from the complexified universe of 'elementary' events $U^c \equiv \{Dr^c\}$. The pair $(U^c, \tau^c_T)$ constitutes *a structured reference background* for the initially considered descriptions $Dr \in U$, which at the start did not exist.

It is clear that the point-representations of a complexified descriptions $Dr^c \in U^c$ are not fit for playing puzzle with them on the basis of semantic attractions by continuity on their borders: the concept of 'borders' has faded away in the abstract representations of the $Dr^c \in U^c$. But we have reached before the liberating conclusion that the requirement of this possibility can be dropped. And the pair $(U^c, \tau^c_T)$ might be found to suffice for insuring the definability of an effective use of the two conditions which have been recognized as essential, namely to permit to build a finite closed whole corresponding to the studied random phenomenon $(\Pi, U)_D$ and to *know* when a replica of this whole had been completed.

But here we stop the development of this introductory example and we go up on the general level of construction where we shall finally outline the whole procedure from *A* to Z.

*A general substitute to the puzzle game with P.* Consider now any random phenomenon *(Π,U)*, *U≡{Dr}*, *r=1,2,....s*. Let *(G,Vr)*, *r=1,2,....s* denote the epistemic referential where the description of this random phenomenon is worked out (*r* is a *global* index each value of which consists of a *set* of definite values of aspect-views *Vg∈Vr*, *all* the aspect-views from the view *Vr* having contributed). Let *[U, τ_T, p_F(τ_T)]* be the factual probability space defined on *U* where *τ_T* is the *total algebra* on *U*. So *τ_T* contains *U* and if we knew the factually true probability distribution *p_F(U)* this would determine also the factually true distribution *p_F(τ_T)* on *τ_T*, via the abstract requirement *p(A∪B)≤p(A)+p(B)* where equality obtains iff *A* and *B* possess no common elementary event.

Quite generally the view *Vr* from the epistemic referential *(G,Vr)*, *r=1,2,....s* where the random phenomenon *(Π,U)*, *U≡{Dr}*, *r=1,2,....s* has been constructed, does *not* involve *all* the aspects that are *directly observable* on an outcome *Dr*. In particular, the spatial qualifications are quasi systematically ignored. While according to the space-time frame-principle (cf. III.5 point *6*) such qualifications are necessarily present in the perception of the outcomes of the procedure *Π*.

Let us form the epistemic referential *(G,V^c)* where the generator *G* is kept unchanged and *V^c ⊃Vr* is a complexified view that contains the initially considered view *V* and furthermore introduces all the observable spatial qualifications. Let *U^c≡{Dr^c}*, with *r^c* a complexified global index including all the directly observable aspects, be the set of all the 'complexifications' of the elementary-event-descriptions *{Dr}≡U*, *r=1,2,....s*, that are conceptually constructible by the help of the view *V^c*. By construction we have *s^c>s*.

The set *U^c* is only an abstract, theoretical universe. Indeed we do *not* know in advance which ones among the conceptually constructible complexified outcomes *Dr^c ∈U^c* do effectively realize when the *initial* experiment *Π* is repeated a very large number of times and the outcomes are examined via the complexified view *V^c*. Indeed the initial random phenomenon *(Π,U)* tied with the descriptions *{Dr}≡U*, *r=1,2,....* to be studied has been constructed under controls tied with only the rougher view *Vr* to which certain conditions of *factual* realization of this or that *Dr^c ∈U^c* qualified via the more complex view *V^c* are likely to have escaped, notwithstanding the conceptual constructability of *Dr^c*.

Consider now the total algebra *τ̃_T* on *U^c*. Like in the introductory example developed before, the *elementary*-event descriptions *Dr∈U* from the initial probability space *[U≡{Dr}, τ_T(U), p_F(U)]* have now migrated into the total algebra of events *τ̃_T* on *U^c*. And there – with respect to the complexified view *V^c* – they appear retroactively as *'simplified' event*-descriptions reduced to exclusively qualifications via the initial aspect-view *Vr∈V^c*. So now, with respect to the new complexified view *V^c*, an outcome *Dr* appears as (theoretically) realizable via a whole set *{Dr^c(r)}*, *r* fixed, *r^c(r)=1,2,....s^c(r)* of elementary-event-descriptions *Dr^c(r)∈U^c* of which the exact *factual* content is not known in advance. But we know by construction that *two sets {Dr^c(r)} and {Dr^c(r')} with r≠r' are mutually exclusive*. Like in the introductory example again, the pair *(U^c, τ̃_T)* is just a medium of conceptual embedment and reference organized for the elementary-event-descriptions *Dr∈U* in order to become able to qualify them in a way more detailed than that which is possible inside the initial epistemic referential *(G,Vr)*.

Let us now construct explicitly the representation space of the complexified view *V^c*. Since each elementary event *Dr^c ∈U^c* is a *physical* entity, according to the *MRC* frame-principle (III.5, point *6*) the complexified view *V^c* necessarily includes some convenient space-frame-aspect-view *V(E)* endowed (in particular) with some definite choice of a length unit that determines the values of the space-coordinates of the directly observable locations of the physical aspect-values *gk* involved by the outcome *Dr^c*.

*We admit that – by the construction of the random phenomenon (Π,U), U≡{Dr}, r=1,2,....s – the set of all the possible space-location coordinates is confined inside a delimited spatial zone Z.* Inside *Z* the complexified elementary events *Dr^c ∈U^c* can be mutually individualized as exactly as one wants, in consequence of the practically unlimited precision that can be assigned to spatial qualifications. Furthermore, we suppose that convenient units have been chosen for also all the semantic axes introduced by all the other aspect-views *Vg∈V^c* that do admit of measurement. Finally, we suppose that on the semantic axes introduced by aspect-views involving *non*-measurable aspects, the values that are taken into consideration have been specified with some definite order, *even if an arbitrary one*.

When all this has been achieved, one can endow the representation space of $V^c$ with a definite *finite* points-grid $\gamma(V^c)$ which covers and *exceeds largely* the spatial zone $Z$ where are confined by construction the factual outcomes $Dr \in U$ of the initial random phenomenon $(\Pi, U)$ to be studied..

By our construction again, *the representation on the grid $\gamma(V^c)$ of any factual realization of a complexified representation $Dr^c \in U^c$ of an outcome $Dr \in U$, is **unique***.

**The algorithm of semantic integration**. From now on an effective identification of the factual numerical probability distribution $\{p_r(U)\}$ can be attempted via considerations entirely similar to those which in IV.1.5 and IV.2 concerned the probabilistic game with the puzzle of the painting $P$.

As many replicas of the points-grid $\gamma(V^c)$ as are necessary can be made use of. Let us start the realizations of the experiment $\Pi$ from the random phenomenon to be studied $(\Pi, U)$. Via the complexified view $V^c$ the first outcome $Dr \in U$ is *observable* as a *complexified* outcome $Dr^c \in U^c$ which, inside $\hat{\tau}_T$, realizes that $Dr \in U$. The representation with respect to the complexified referential $(G, V^c)$ of this first observed outcome $Dr^c \in U^c$ falls on a uniquely corresponding point from the *first* replica of the points-grid $\gamma(V^c)$, which we denote by $\gamma 1 (V^c)$.

While $N$ increases, any new outcome will be represented:

- *either* on the first replica $\gamma 1(V^c)$ of the grid $\gamma(V^c)$, at the uniquely corresponding place indicated by the newly observed value of the complexified index $r^c$, *if* this place is not yet occupied on $\gamma 1(V^c)$;

- *or*, if on $\gamma 1(V^c)$ the place indicated by the newly observed value of the complexified index $r^c$ is already occupied, the new outcome will be represented on the *nearest* replica $\gamma k(V^c)$ of $\gamma(V^c)$ where that place is still available, with $k=1,2,....K$ an integer that labels the already introduced replicas of $\gamma(V^c)$ and $K$ also a finite integer.

The total number of points on one replica of the grid $\gamma(V^c)$ is finite by construction and any factual outcome of the experiment $\Pi$ is represented with respect to $V^c$ by a *unique* place on $\gamma(V^c)$. So the way of acting specified above entails the progressive emergence on $\gamma 1(V^c)$ of a cloud of points for which an evolving dotted *delimitation* acquires definition because by construction $\gamma(V^c)$ covers largely the zone $Z$ where factual outcomes $Dr \in U$ can emerge: in *this* sense we can speak of *the emergence on $\gamma 1(V^c)$ of a first representation $\phi 1$ of a still unknown but confined 'points-form' $\phi$*. Moreover – in the same time – also a sequence of other emergent more and more *unachieved* reproductions of this same points-form $\phi$ will have become observable on subsequently introduced replicas $\gamma 2(V^c)$ $\gamma 3(V^c),...$ $\gamma k(V^c)...,\gamma K(V^c)$ of the points-grid $\gamma(V^c)$.

Sooner or later but in a *finite* time because everything is finite by construction, the first replica $\phi 1$ of $\phi$ will be observed to *cease to evolve* while the number $N$ of realizations of the procedure $\Pi$ continues to grow and to nourish the growths of the subsequent emergent points-forms labeled by $k=2, 3,.... K$. Thereby we shall *know* that $\phi 1$ is quasi certainly *completed*.

Here we stop to make immediately a basic remark concerning the locution 'quasi certainly'. We employ this locution because nothing can exclude absolutely that never at some future time, while $N$ increases, an outcome $Dr^c$ will realize that finds its place still free on $\gamma 1(V^c)$, so which adds a new element to the points-form $\phi$. All the following considerations are marked by this permanently pending possibility. But *any* natural law is marked by similar possibilities. And even a *radical* future modification keeps always possible, for a factual probability distribution as well as for any other sort of natural law. Both distributions of probability and natural laws asserted as certainties are but local conceptual constructs founded on a postulation of unchanged 'conditions' (MMS [2002] pp. 291-303). A certainly definitive, an absolutely stable factual truth can never be constructed conceptually. Only syntactic 'truths' can be absolutely stable[30]. But these are conclusions of deductions, truths of logical inner coherence, not observational factual truths. (Moreover their sort of stability also, though in contradistinction to factual truth it can last indefinitely, nevertheless is indelibly relative, namely relative to the necessarily restrictive syntax where they have been established).

We now resume the interrupted development. Suppose that we have continued to increase the number $N$ of realizations of the experiment $\Pi$ and that this has produced, say, $K$ integrated point-forms *that are all identical with* $\phi 1$, as well as a new sequence of less and less achieved emerging point-forms. We denote by $\phi(K)$ *any* one among these $K$ mutually identical point-forms and we call it *the K-variant of the unknown point-form* $\phi$. On $\phi(K)$, just like on the integrated puzzle of the picture $P$, we can *count* the total number $n^c_{\phi,T}$ of all the *complexified* outcomes $Dr^c_{\phi}$ that have been *factually* realized there[31]. And we can equally *count*, for any *given* description $Dr \in U$ reconsidered as an event from the algebra of events $\hat{\tau}_T$, the number $n^c_{\phi(K)}(r)$ of realizations *inside* $\phi(K)$ of *that Dr*, via this or that complexified factually realized outcome from the set of all the factual outcomes $\{Dr^c_{\phi(K)}(r)\}$ with $r$ fixed and $r^c_{\phi(K)}(r)=1,2,....s^c_{\phi(K)}(r)$. For the cardinal $s^c_{\phi(K)}(r)$ of this *factually* realized set we can now write: $s^c_{\phi(K)}(r) \equiv_F n^c_{\phi(K)}(r)$ (the sign '$\equiv_F$' is to be read: factually identical to): this cardinal has acquired a meaning with respect to $\phi(K)$.

By construction a complexified description $Dr^c$ that is element of $\{Dr^c_{\phi(K)}(r)\}$ cannot be also element of a set $\{Dr^c_{\phi(K)}(r')\}$ where $r' \neq r$. Equally by construction we have $n_{\phi(K),T}=\Sigma_r n^c_{\phi(K)}(r)$ and $\Sigma_r n_{\phi(K)}(r)/n_{\phi(K),T}=1$, $r=1,2,....s$. So – on the basis of considerations that are strictly analogous to those from IV.2 concerning the probabilistic game with the puzzle of the picture $P$, but are quite *generally* founded this time – we can assert that via a probability game with the elements $Dr^c_{\phi(K)}$ of the finite and confined 'form' $\phi(K)$, the factual numerical definition of the probability of an event $Dr$ is found to be the set of **rational** and *factually* registered numbers:

$$\{p_F(r,K)\} \equiv \{n_{\phi(K)}(r)/n_{\phi(K),T}\}, \quad r=1,2,....s \qquad (4')$$

Simulations by computer might permit to organize rather rapidly all the substratum for the estimations from *(4')*.

The procedure that led to the definition *(4')* will be called *the algorithm of semantic integration of the factual numerical probability distribution to be asserted on the universe U generated by the random phenomenon (Π,U).*

This algorithm is purely factual. The possibly unending confrontations between an a priori assertion of a uniform distribution of factual probabilities and an a posteriori verification of this assertion by measurements of relative frequencies, has been dissolved in a unique sequence of exclusively factual operations.

But is this algorithm also clearly effective? *How* large should the integer $K$ be in order to be authorized to assert the distribution *(4')*? As soon as one seeks the answer to this question, the definitions of the involved concepts lead to a radical *simplification* of the algorithm outlined above:

What happens if, while $N$ is still increased, a new point-event $Dr^c$ is observed of which the place is still *un*occupied on $\gamma 1(V^c)$, even though before the content of $\gamma 1(V^c)$ had remained unchanged during an already long succession of repetitions of $\Pi$? Well, this would reveal that the point-event $Dr^c \in U^c$ is *factually* possible, in other words, that its factual probability – whatever it is – is *not null*. But the statement of a non-null probability of that new $Dr^c$ is equivalent to asserting that if the number $N$ of repetitions of the experiment $\Pi$ were very much increased, $Dr^c$ would quasi *certainly* reappear from time to time, so that, progressively, it would occupy its place on *all* the already $K$ constructed

---

replicas identically denoted $\phi(K)$. This amounts to a retroactive modification of the concept *'$\phi(K)$'*. But we were prepared for this possibility: precisely because we have conceived it we have specified *the relativity to K* of the succession $\phi1$, $\phi2$, $\phi3$, ,... $\phi k$,... $\phi K$ of the identical points-forms which we had obtained.

On the basis of this reasoning *any* event $Dr^c$ which still finds its place free on $\phi1$, even after a very long period of apparent saturation, can be immediately reproduced also all the other possible replicas of $\phi$. So, as soon as we are already endowed with a first replica $\phi1$ of $\phi$ that behaves for a time as it were saturated, we are in fact endowed *during this time* with an *arbitrarily long* sequence $\phi2$, $\phi3$, ,.....$\phi k$,... of mutually identical point-forms. This is entailed by the fact that, in contradistinction to the example of the probabilistic game with puzzle of the painting $P$, in the general case treated here the presupposed form $\phi$ *itself* is unknown at the start, not only the factual distribution of probabilities tied with it. So, by absence of a definite reference, we never know certainly whether a replica of the posited form $\phi$ is achieved, or not. However, the mere postulation of the existence of a form $\phi$ together with the deliberate construction of the conditions for observable manifestations of this existence – namely a stability of the content of $\phi1$ while a whole succession $\phi2$, $\phi3$, ,..... $\phi k$ ,... $\phi K$ of subsequent replicas of $\phi$ emerges and develops – induce already a peculiar sort of *quantification* of the number $N$ of achieved repetitions of $\Pi$: say, a $K$-quantification. For these manifestations of the existence of a form $\phi$, though relative to the number $K$, act throughout the process of emergence of all the points-forms $\phi(K)$ like a sort of punctuation that breaks the amorphous flux of the uniformly increasing number $N$ of realizations of $\Pi$ from the theorem of large numbers, in a sequence of mutually separated entities. Precisely here lies the importance of being able to associate with the random phenomenon $(\Pi,U)$, a concept of a finite and delimited whole. Even if we cannot know with rigorous certainty whether this punctuation is like a full-stop or like a semi-colon, so whether yes or not the content of outcomes $Dr^c$ already registered inside all the replicas $\phi(K)$ exhausts the content of the unknown form $\phi$, nevertheless the mentioned quantification *marks* – by the parameter $k$ – the places reachable by a finite number of repetitions of $\Pi$ where we can decide to *cease* the research and to draw already a conclusion relative to the number $K$, namely the conclusion defined by *(4')*.

In *this* sense the algorithm of semantic integration of *(4') is* effective. And one cannot hope more:

*As long as we stay inside the framework of the probabilistic conceptualization, effectiveness tied with rigorous certainty is just* **contradiction**.

But of course, the exact relation between the sort of effectiveness entailed by the algorithm of semantic integration, and the theorem of large numbers, must be examined more thoroughly. Just below however we shall first draw attention to the main conceptual consequences of the algorithm of semantic integration.

**Significance of the 'existence' of a factual probability distribution**. The algorithm of semantic integration involves radically new and quite definite *non*-mathematical *meanings* for the notion of a factual distribution of numerical probabilities on the universe $U$ generated by the random phenomenon $(\Pi,U)$:

- The 'existence' of a factual probability distribution – the *mere* existence whatever the content – is equivalent to the assertion of a progressive process of emergence and saturation, on the grid $\chi(V^c)$ constructed for $(\Pi,U)$, of a points-form $\phi$ made up by the representation-points of all the *factually* possible complexified elementary-event-descriptions $Dr^c$ from the reference-universe $U^c$.

- The numerical distribution of probabilities itself expresses the *way* in which, when the points-form $\phi$ has reached a quasi certain stability inside $K$ replicas of $\phi$, the probabilities of the event-descriptions $Dr \in U$ are defined inside this form *relatively* to $K$, as well, of course, as relatively also to all the other involved restrictions.

*Individual description versus probabilistic description*. The meanings stated above identify the specificities of a probabilistic description inside the general category of descriptions of physical entities: According to *MRC* any communicable and consensual knowledge is description, so precisely a points-form of space-time-and-aspect-values endowed with some invariance, either individual or probabilistic (III.5, points *5, 6, 11*).

When the description is individual, the corresponding form of space-time-aspect-values, usually; is immediately 'understood', it carries an obvious 'significance'.

But when the description is tied with a random phenomenon *(Π,U), U≡{Dr}, r=1,2,....s* the involved physical situation in general escapes a direct and integrated human perceptibility, because it involves features too tiny, or too large, or partly hidden, and anyhow too complex to be immediately put together inside one organized structure. Moreover, in the fragments of the involved integral physical situation which *are* directly perceived, certain spatial features or more generally certain space-time features[32] (distances, angles, relative durations, etc.), are occulted; they are filtered out more or less conventionally or arbitrarily. Only certain isolated indications stemming from this integral structure are directly observable *and noticed*. What has been denoted *Dr∈U* represents only such indications; and the lacunae with respect to the global physical organization which is at work, destroy the *intelligibility*. They destroy even the capacity to merely *imagine* the existence of the involved factual 'whole'. The relative frequencies of the outcomes of the watched 'elementary' events *Dr∈U*, by their observable tendency toward *stability* when the number *N* of repetitions of the procedure *Π* increases, construct progressively – by parceled random touches – a purely numerical and radically cryptic representation of the unknown physical whole from which they stem. They generate a sort of coded, random and approximate asymptotic 'reading' of this whole, which conveys only impoverished and pulverized signals from it. And even when, a posteriori, these relative frequencies are considered simultaneously for all the events *Dr* and with their stabilized values, the factual probability distribution *{p_F(r,K), r=1,2,....s* which emerges still yields merely a meaningless numerical expression of the now fully accomplished, but impoverished and randomized process of reading in cryptic terms the unknown physical whole which generated the distribution.

Here comes in the remarkable role of the *'complexifications' Dr^c(r) ∈U^c* constructed for the directly perceived event-descriptions *Dr∈U*. These complexifications, by embedment and reference, permit to accede a more or less conventional representation of a global form *φ* associable with *(Π,U)*. Because of the various conventions involved in its construction, this global form is only an element of a whole class of forms equally possible with respect to the data *(Π,U), U≡{Dr}, r=1,2,....s*. The representations from this class *still* offer an only *coded* integrated description of the whole conceived to be at work. Indeed the involved qualia are only represented, they are not perceived, and – as it is requires by the conventional character of certain representational choices – any specifications of the ways in which do realize in fact the juxtapositions of the complexified descriptions *Dr^c∈U^c* represented by the points from *φ*, are entirely lacking. But in spite of all these lacks, the mutual *individualization* by the complexified outcomes *Dr^c(r)∈U^c* of 'one same' *Dr∈U* and the possibility to count of these outcomes inside a finite and closed whole, suffice for investing with intelligibility the notion of probability of an event-description *Dr∈U*.

**On the truth of the points-form *φ*.** The considerations made above show that it would be meaningless to ask whether the integrated form *φ* is 'true'. It is just a conceptual *construct* achieved with the aim to work out a factual definition for the numerical probability distribution to be asserted on the universe *U* generated by a random phenomenon *(Π,U)*. As for this definition itself, it is true in so far that it can be factually produced and 'verified'.

This brings us back to the necessity of a thorough examination of the logical compatibility between the definition *(4')* and the formal concept of probability and the theorem of large numbers.

---

[32] In all the considerations from this work the role played by spatial features can be extended to space-time features.

*Construction of a factual-formal variant of the theorem of large numbers*

Consider again the expression *(2)* of the theorem of large numbers

$$\forall r, \quad \forall (\varepsilon, \delta), \quad \exists N_0 : \quad \forall (N \geq N_0) \quad \Rightarrow \quad \boldsymbol{P}[(\,/n(e_r)/N - p(e_r)\,/) \leq \varepsilon] \geq (1 - \delta) \qquad (2)$$

This expression has been proved inside the syntactic theory of probabilities. The abstract concept of a probability $p(e_r)$ involved in *(2)* amounts to:

*(a)* Presupposition of the *existence* of a numerically *non*-specified mathematical limit denoted $p(e_r)$ toward which the measured relative frequencies $n(e_r)/N$ are asserted to converge and *definition* of the 'probability of $e_r$' as precisely that limit, for any fixed value of the index $r$ ;

*(b)* Specification of exclusively the well known *general* abstract structure recalled in the note 3 assigned to the whole set of mathematical limits *{p(e_r)}*, *r=1,2...s*, left *void* of any specification of a distribution of numerical values characteristic of this or that factual probabilistic situation.

The same holds for $\boldsymbol{P}$, mutatis mutandis.

Both *(a)* and *(b)* express a non effective and semantically void point of view, deliberately made fit for the role of insuring maximal generality as a mathematical receptacle for any factual probabilistic data. And there is no other semantic content in the abstract concepts of probability of an event and of probability measure on a universe of events.

Consider now the whole equation *(2)*. It represents the syntactic evolution of the relation between the mathematical limit $p(e_r)$ – initially void of specification of a numerical value – and the relative frequency $n(e_r)/N$, while the integer $N$ that counts the accomplished repetitions of the considered experiment $\varPi$ tends toward infinity by uniform ordered steps of one unit integers *1* which *progressively inject a numerical value into the sign $p(e_r)$.*

The uniform progression of $N$ extends over the whole abstract interval from *1* to $\infty$ and it runs through this extension in a way that is *blind* with respect to *semantic* peaks of 'significance' of the ratio $n(e_r)/N$ with respect to the definability of a factual probability $p_{\phi F}(Dr,K) \equiv n_{\phi(K)}(r)/n_{\phi(K),T}$. Furthermore, it does not take into account factual human impossibilities concerning the future; correlatively it does not distinguish between a priori hypothetical factual assertions and a posteriori factual findings. It just goes on toward $\infty$ by mutually equal steps. The fluctuations in the way of evolving of the value of the ratio $n(e_r)/N$, are free of any regulation. The syntactic way of fighting these fluctuations inside the framework of the equation *(2)* is only their *méta-probable* confinement imposed by a pair *(ε,δ)* of arbitrarily small real numbers.

Let us compare with the quantifying procedure of semantic integration where mainly the semantic contents and their evolution are watched. While the number of successive realizations of $\varPi$ is increased from *1* to $n_{\phi(K),T}$ the variable $N$ evolves inside a domain of 'absence of a full significance' *with respect* to the concept of probability $p_{\phi F}(Dr,K) \equiv n_{\phi(K)}(r)/n_{\phi(K),T}$ with $r=1,2,....s$. Indeed with $1 \leq N \leq n_{\phi(K)T}$ the points-form $\phi(K)$ is not yet achieved, so the counting that determines the probabilities from the factual distribution *(4')* cannot be assumed to be fully significant. The same holds for $n_{\phi(K),T} \leq N \leq 2n_{\phi(K)T}$, $2n_{\phi(K),T} \leq N \leq 3n_{\phi(K),T}$ ...... So with respect to the semantic considerations and to definability of the probabilities $p_{\phi(K)F}(Dr) \equiv n_{\phi(K)}(r)/n_{\phi(K),T}$, $r=1,2,....s$ from *(4')* there are peaks $N=kn_{\phi(K),T}$ with $k=1, 2,.....K$ of full significance outside which this relative significance is gradually increasing or decreasing with a period $\Delta N=n_{\phi(K),T}$.

But from a purely numerical point of view, the algorithm, just like the equation *(2)*, introduces a uniform and ordered progression of the integer $N$. Meanwhile however this algorithm also takes into account simultaneously all the semantic contents of *all* the outcomes $Dr^c(r)$, $r=1,2,....s$ and on the basis of *these* it selects the place of each outcome $Dr^c$ on this or that replica of the grid $\gamma(V^c)$ quite *independently of the order of its emergence inside the sequence of 1,2,3.....N of repetitions of the experiment $\varPi$*. In this way, by semantically founded locations of the outcomes, the algorithm separates from inside the steadily increasing sequence of repetitions of the procedure $\varPi$, accomplished *K*-quanta of significance with respect to the aim of defining the set of factual probabilities $p_{\phi(K),F}(Dr) \equiv n_{\phi(K)}(r)/n_{\phi(K)T}$, $r=1,2,....s$.

These semantically regulated organizations of the successively registered results $Dr^c$ produced by the uniform progression of the number $N$, are what absorbs and dissolves into a unique and entirely factual approach, the necessity of possibly never ending [a priori-a posteriori] dialogs, which up to now could not be avoided.

Let us now examine closely the formal compatibility between the algorithm of semantic integration and the theorem of large numbers *(2)*: We shall attempt to embed the algorithm of semantic integration, into the abstract framework of the equation *(2)*.

We introduce the identification of terms $e_r \equiv Dr$ and from now, for simplicity, we renote *Dr* by *r*. We *fix* a given value *r*, as in the equation *(2)*. Furthermore in the absolute difference from the equation *(2)* we substitute to the *unspecified* but *real* numerical value $p(e_r)$, the *rational* number $n_{\phi(K)}(r)/n_{\phi(K),T}$ specified by *(4′)*. Thereby $p(e_r)$ acquires a factual specification of its numerical value which in *(2)* is lacking, while the rational number $n_{\phi(K)}(r)/n_{\phi(K),T}$ acquires the *syntactic* definition from *(2)* for the probability of an event *Dr* which for the factual probability $p_F(r,K)$ was lacking in *(4′)*, namely its definition as the mathematical limit toward which the classical factual ratio $n(r)/N$ must tend when *N* increases toward infinity via its semantically blind evolution. So with $k$ =*1, 2,...K* and *K* indefinitely increasable we have:

$[p(e_r) \equiv p_F(r,K)] =$
$= \Sigma_k n_{\phi(K)}(r)/\Sigma_k n_{\phi(K),T} = K\,n_{\phi(K)}(r)/K\,n_{\phi(K),T} = n_{\phi(K)}(r)/n_{\phi(K),T} = lim_K.N \to \infty\ (n(r)/N)$      *(5)*

The symbol $lim_K\,N \to \infty$ means $lim.N \to \infty (n(r)/N)$ relatively to the hypothesis that the points-form $\phi(K)$ exhausts the unknown form $\phi$. We furthermore introduce the new notations:

$\qquad N=N(K)=Kn_{\phi(K),T}+N'$,   $n(r)= Kn_{\phi(K)}(r)+n'(r)$            *(6)*

where the term $Kn_{\phi(K),T}$ represents the number of unit steps from *N* which have got consumed in the construction of all the mutually identical *K* points-forms $\phi(K)$ and so have introduced $Kn_{\phi(K)}(r)$ realizations of the event *Dr*. The term $N'$ represents the supplementary number of unit steps from *N* which have got consumed in the construction of emergent but not yet achieved drafts of the unknown form $\phi$ and so have introduced some unknown number $n'(r)$ of realizations of *Dr* that are not yet 'significant' with respect to the definition *(4′)* of $p_F(r,K)$.

With the notations introduced above, the absolute difference from *(2)* becomes

$/n(r)/N(K) - p_F(r,K) / = / [(K\,n_{\phi(K)}(r) + n'(r))/(K\,n_{\phi(K),T} + N')] - (n_{\phi(K)}(r)/n_{\phi(K),T}) /$     *(7)*

So we finally obtain the following 'factualized' and numerically specified variant of the theorem of large numbers:

$\forall r,\ \ \forall(\varepsilon,\ \delta),\ \forall K,\ \ \ \ \exists N_0:\ \ \forall((N(K) \geq N_0),\ N(K)=K\,n_{\phi(K),T} + N'\ \ \Rightarrow$
$\Rightarrow \textbf{P}[\ /\,[(K\,n_{\phi(K)}(r) + n'(r))/(K\,n_{\phi(K),T} + N')] - (n_{\phi(K)}(r)/n_{\phi(K),T})\ /\ \leq \varepsilon\,]\geq (1 - \delta)$    *(2′)*

By trivial transformations the absolute difference *(7)* involved by *(2′)* becomes

$\qquad /\,(\,n_{\phi(K),T}\ n'(r) - n_{\phi(K)}(r)\ N')/(K(n_{\phi(K),T})^2 + n_{\phi(K),T}\,N')\ /$             *(7′)*

On *(7′)* it can be explicitly seen that when the number *K* (so *N(K)*) is increased indefinitely, the difference from *(7)* and *(2′)* tends toward zero because: $n_{\phi(K),T}$ and, $n_{\phi(K)}(r)$ are quasi certainly constant; $n'(r)$ and $N'$ can be considered to keep constant *in the mean* even though in general they fluctuate when a passage occurs from a given value *K* to the value *K+1*, because the general physical conditions are invariant with respect to the value of $K$ that marks such a passage; while *K increases indefinitely and $n_{\phi(K),T}$ is bigger than both $n_{\phi(K)}(r)$ and $n'(r)$.*

The form *(2')* of the theorem of large numbers has been obtained by just injecting into the expression *(2)* the factual structure *(6)* of the numbers *N, n(r)* defined for these numbers by the algorithm of semantic integration, as well as the numerical factual value $p_F(r,K)$ assigned by this algorithm for the abstract probability *p(Dr)* defined syntactically as *lim.N→∞(n(Dr)/N)*.

The substitution *(5)* to the unknown but fixed number $p(e_r)$, of the possibly evolving and 'quantified' number $p_F(r,K)$, is the core of a *change* in the general view expressed by *(2')*, with respect to the view expressed by *(2)*: the number $p_F(r,K)$ is no more regarded as a fixed limit, but as a possibly evolving term (though for $K$ of the order of several units the evolution of the value of $p_F(r,K)$ is already unlikely). But this change of view generates no formal contradiction inside the very tolerant $(\varepsilon,\delta)$-approximate and (meta)-*probabilistic* (via **P**) syntactic framework of the theorem of large numbers: this framework remains blind with respect to the mentioned change.

On the other hand it permits to complete this framework by semantically founded numerical specifications which in *(2)* are left free.

The result *(2')* is now a *non circular* syntactic-semantic proposition, in the following sense. The definition *(4')* of a factual probability distribution $p_F(r,K)=[n_{\phi(K)}(r)/n_{\phi(K),T}$ changes *nothing* in the numerical values of the ratios *n(r)/N* from (2) and in the evolution of these[33]. Now, a definition of the type *(4')* which does not interfere with the purely numerical aspects of the relative frequencies of the events to which this definition refers, can be constructed for *any* probability. So in particular, a definition of type *(4')* can be constructed *also* for the meta-probability **P** from *(2)* (in that case the considered 'events' are the numerical values acquired by the absolute difference from *(2')* for each choice of a set of values *(r, (ε, δ), K)*). This general possibility entails that the form *(2')* ceases to define a probability $p(e_r)$ by use of another probability **P**, since the probabilities $p(e_r)$ and **P** possess *own* and mutually independent definitions which do not alter the relative frequencies tied with them, respectively. So *(2')* asserts only relations between the probabilities $p_F(r,K)$, **P**, and the involved – respective – relative frequencies. Namely, the new, factual, *semantic-syntactic* and non circular proposition *(2')* asserts – as the expressions *(7)* and *(7')* show – that the involved absolute difference concerns only the 'non-significant excesses' *n'(r)* and *N'* from the ratio *n(r)/N)*, with respect to the peak of exact definability of the factual probability $p_F(r,K)$ by use of the algorithm of semantic integration.

As for the proof of the whole theorem of large numbers, it insures that the chosen pair of real numbers $(\varepsilon,\delta)$ and the meta-probability **P** guarantee that the value of the absolute difference from *(7)* and *(2')* – calculated for the hypothetical and possibly *variable* numerical value *(4')* assigned to the probability $p_F(r,K)$, which in *(2)* is **not** *defined* – tends toward *0* in probability when $K$ (so *N(K)*) is indefinitely increased (indeed since rational numbers are real numbers the substitution *(5)* amounts to considering a particular class of mathematical cases). Thereby the theorem of large numbers *takes in charge* both the a priori possible fluctuations of what we have called 'the quasi certain completeness' of the points-forms *$\phi(K)$*, and the existence, in general, of fluctuations of the numbers *n'(r)* and *N'* while $K$ is increased.

We conclude that there *is* logical consistency between the algorithm of semantic integration and the theorem of large numbers; and that moreover there is synergy between them. But in fact, for cases where the 'quantum' $n_{\phi(K),T}$ is large, the algorithm of semantic integration permits far more rapid and more informed estimations than those which can be drawn from *(2')*, of the factual probability distribution to be asserted on the universe $U$ of outcomes generated by a random phenomenon *(Π,U)*.

---

[33] We have already stressed, and we repeat, that the translation of this ratios in our terms – $n(r)/N=[(Kn_{\phi(K)}(r)+n'(r))/(K\,n_{\phi(K),T}+N')]$ – point exclusively toward the semantic choices of the locations of the outcomes $Dr^c$ of the $N$ successive repetitions of the experiment $\Pi$, they change nothing in the pure numeration of these repetitions and of their outcomes.

*Conclusion*

The algorithm of semantic integration solves Kolmogorov's aporia.

It does not solve it by a deductive approach, but by a constructive procedure which is formally compatible with the theorem of large numbers and admits integration in it.

It seems difficult to advance further in the attempt at associating in a common coherent view, a syntactic, fully deductive structure like the theorem of large numbers, and on the other hand a factual procedure achieved inside a qualitative method of conceptualization. Insofar that one accepts that the attempt developed here offers a satisfactory association of this kind, this attempt organizes the whole classical concept of 'probability'.

## V. ON 'PROBABILITIES' IN MICROPHYSICS

The algorithm of semantic integration *cannot* be applied to the random phenomena tied with microstates. Indeed for these, 'complexifications' $Dr^c$ of the observed results cannot be constructed. The results of measurements on a microstate quasi generally consist of *exclusively* space-time locations of observable marks that do not produce in the observer's mind any *qualia* carrying a definite semantic content. The mere space-time *location* of these marks is pre-coded in terms of values of some 'quantity' assigned to a microstate, on the basis of *hidden, 'illegal', more or less implicit and non consensual, but nevertheless utilized* **models** *of a microstate and of its mechanical characters*.

These mere space-time locations of observable marks, void of own direct significance, cannot in any way be observably expanded in more complex events.

This – surprisingly – entails that what is quite unanimously called the 'essential' or 'primordial' *probability laws* established for microstates (cf. MMS [2009]), in fact are merely *statistical primordial distributions* for which the existence of a mathematical limit can be, at most, just asserted and – more or less – checked by an observed *tendency*, in a way that *never* is strictly effective.

As for the possibility to determine the factual quantum mechanical probability law to be asserted in any given probabilistic situation concerning microstates, by *deriving* this law from the mathematical representations that constitute the theory, its generality seems highly problematic (cf. note 14). So, *in general*, with respect to a microstate, one is reduced to only indications of a tendency toward convergence obtained by *measurements* of relative frequencies performed in physical conditions consistent with the mathematical representations.

This conclusion, which probably is very personal, is here *submitted* to the physicists for discussion: It might come out that the concept of probability is constructible in an effective way only inside the domain of *classical* thinking where the entities-to-be-qualified are directly perceptible *with qualia different from* **exclusively** *their space-time locations* and where *models* founded on these direct more complex perceptions can be built and expressed in terms of 'objects' (MMS [2006] pp. 118-127)[34].

## VI. GENERAL CONCLUSION

In the present phase of the scientific thinking the concepts of statistics and of probability require a thorough reconstruction if it is wanted to bring them to cover in an exhaustive and coherent way both the classical thinking and nowadays microphysics.

This work has been intended as an attempt on this direction.

At the same time it has been conceived as also an illustration of the way in which the method of relativized conceptualization can be made use of in order to clarify concepts and questions and to construct solutions to the questions.

---

[34] Here these are still only *conjectures* formulated in order to be discussed.

---

[35] At pp. 119-122, instead of *"{$\mu_{iq}$, j=1,2,...m)}$\leftrightarrow Xj$"* please read – everywhere where this comes in – *"{$\mu_{iq}$, k=1,2,...m)}$\leftrightarrow Xj$* (with *q* an index of order that labels the considered 'interaction). The same holds for the reference MMS [2009B].